\documentclass[article]{jfm}

\usepackage{lineno,hyperref}
\usepackage{color}
\usepackage{graphicx}
\usepackage{subfig}
\usepackage{epstopdf}
\usepackage{natbib}
\hypersetup{
    colorlinks = true,
    urlcolor   = blue,
    citecolor  = black,
}
\usepackage{comment}
\usepackage{xcolor}

\usepackage{arydshln}
\usepackage{tikz}
\DeclareRobustCommand\full  {\tikz[baseline=-0.6ex]\draw[thick] (0,0)--(0.5,0);}

\DeclareRobustCommand\dashed{\tikz[baseline=-0.6ex]\draw[thick,dashed] (0,0)--(0.54,0);}
\DeclareRobustCommand\dashdotted {\tikz[baseline=-0.6ex]\draw[thick,dash dot dot] (0,0)--(0.5,0);}

\newcommand{\RomanNumeralCaps}[1]

\renewcommand{\figurename}{Fig.}

\newcommand{\ddr}[1]{\ensuremath{\frac{\partial #1}{\partial r}}}
\newcommand{\dddr}[1]{\ensuremath{\frac{\partial^2 #1}{\partial r^2}}}

\newcommand{\vjt}[1]{{\color{blue}#1}}

\modulolinenumbers[5]

\title{\vjt{\begin{center}Linear dynamics of over-expanded annular supersonic jets\end{center}}}

\author{\begin{center}Vincent Jaunet\aff{1}, Guillaume Lehnasch\aff{1}\end{center}}

\affiliation{\begin{center} \aff{1}ISAE-ENSMA,
	Institut PPrime, Universit\'e de Poitiers, UPR-3346 CNRS\\
	1 Avenue Clement Ader, 86360 Chasseneuil-du-Poitou, France \end{center}}

\pubyear{1}
\volume{1}
\pagerange{1}
\date{?; revised ?; accepted ?. - To be entered by editorial office}

\begin{document}

\maketitle

\begin{abstract}
This article delves into the dynamics of inviscid annular supersonic jets, akin to those exiting
converging-diverging nozzles in over-expanded regimes. It focuses on the first azimuthal Fourier mode
of flow fluctuations and examines their behavior with varying mixing layer parameters and expansion
regimes. The study reveals that two unstable Kelvin-Helmholtz waves exist in all cases, with the
outer layer wave being more unstable due to velocity gradient differences. The inner layer wave is
more sensitive to base flow changes and extends beyond the jet, potentially contributing to nozzle
resonances. The article also investigates upstream propagating guided-jet modes, which are found to
be robust and not highly sensitive to base flow changes, making them essential for understanding jet
dynamics. A simplified model is used to obtain ideal but base flows with realistic shape to study
varying nozzle pressure ratios (NPR) effects on the dynamics of the waves supported by the jet.
\end{abstract}



\nolinenumbers


\section{Introduction}
\begin{figure}
  \centering
  \includegraphics[trim={0cm 0cm 0cm 0cm},clip, width=0.8\textwidth]
  {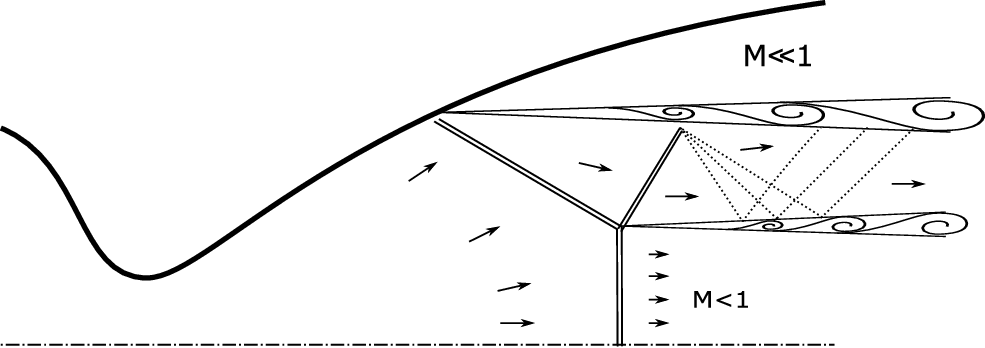}
  \caption{Illustration of TIC nozzle jet in Free Shock Separation (FSS) regime.}
  \label{fig_tic}
\end{figure}

Thanks to the entry of new participants in the aerospace market, the interest in space access has
recently been rekindled. This active competition, occurring in parallel with global environmental
concerns, compels industries to develop launch systems that are more robust, efficient, and
cost-effective for deploying satellites into specific orbits. To mitigate expenses, reusable
launchers and boosters have been designed, enabling their return and landing on the Earth's
surface. During the descent and landing phases, a comprehensive understanding of the aerodynamics of
the jet plume, exhausting at supersonic speed from the convergent-divergent nozzle, is imperative to
avoid unintended mechanical stresses arising from uncontrolled pressure fluctuations. The startup
phase of the engines also confronts
side-loads, which has motivated extensive research
(see
\cite{Nave1973,Schmucker1973-1,Schmucker1973-2,Schmucker1973-3,Schmucker1974flow,Dumnov1996,Deck2004,deck2009delayed}
and reference therein for a subset of studies). Despite the fact such issues may become
even more critical during landing, where the jet may impinge on a flat surface, comprehensive and
quantitative models for predicting unsteady pressure forces are still to be developed.\\
Resonances in jet flows originating from "truncated ideal contour" (TIC) nozzles, and operating
within the free separation regime, have been observed within very narrow over-expansion ratio ranges
\citep{Baars2012, jaunet2017wall, martelli2020flowdynamics}. Recent investigations have demonstrated
that the associated pressure disturbances are likely to generate significant lateral, off-axis, forces that
could lead to structural damage or even flight control problems
\citep{bakulu2021jet}. \vjt{Despite the research efforts, no consensus on the origin of these
  resonances have yet emerged and among the various possible explanations the role of vortex
shed by the Mach disk \citep{martelli2020flowdynamics}, the effect of the separated mixing layer
impingement in the nozzle lip \citep{tarsia2023impinging} or an internal screech feedback loop
\citep{jaunet2017wall} were hypothesized. The seek for the necessary ingredient for a feedback loop
to be sustained in these atypical flows is one of the motivations for this study.}\\
In these particular flow regimes, the presence of Mach disks in the flow decelerates the central core of the
jet, while a higher-velocity annular flow envelops this slower core, as depicted in figure
\ref{fig_tic}. The mean flow thus comprises two co-annular jets, separated by a mixing layer
emanating from the separation point and a slipline originating from the Mach disk triple point. The
dynamics of these flow conditions have been shown to play a crucial role in the resonance mechanism
and warrant more in-depth investigation to gain insights into the resonance process
\citep{bakulu2021jet}.\\
Axisymmetric bulk velocity profile jets (single jets) have garnered significant attention within the
scientific community, \vjt{especially to} predict their acoustics \citep{jordan2013wave}.
Hence, substantial efforts have been dedicated to modeling the dynamics of these jets, employing
linearized models, with the well-known $\tanh$ profile serving as a proxy for the jet's base flow
\citep{michalke1984survey}. Supersonic jet flows have also been extensively studied, and numerous
dynamical models have been developed to understand their dynamics \citep{tam1989three}, noise
characteristics \citep{tam1972noise,tam1991broadband}, and resonances (please refer to \cite{edgington2019aeroacoustic} for a recent review).\\
Coaxial jets have also been a focal point for the scientific community, particularly during the late
1970s when it was discovered that they can exhibit reduced noise levels compared to their equivalent
single jet counterparts under optimal operating conditions \citep{dosanjh1969noise,
  dosanjh1971reduction, yu1971noise}. Extensive experimental studies have shown that supersonic
coaxial jets operated with a faster external stream, denoted as inverted velocity profile (IVP)
jets, can lower the overall jet noise due to a decrease in shock-associated noise. To elucidate this
phenomenon, a series of experimental and theoretical investigations ensued \citep{tanna1985shockI,
  tam1985shockII, tam1985shockIII}.\\
In their experimental study, \cite{tanna1979noise} examined shock-free IVP supersonic jet
noise. They observed that IVP jets produce louder high-frequency noise at all angles and quieter
low-frequency noise near the jet exit axis. These variations were found to be more pronounced when
the velocity ratio of the two streams exceeded 1. The authors concluded that the rapid decay of the
maximum mean velocity in IVP jets is a significant factor contributing to the noise reduction in IVP
jets compared to single jets. Linear stability studies were subsequently employed to explain the
observed trends by \cite{bhat1993effect}, and later, by \citep{dahl1997noiseI, dahl1997noiseII,
  dahl1997noiseIII}. In particular, the analysis in \cite{dahl1997noiseIII} yielded critical
insights into the dynamics of inverted velocity profile (IVP) jets. The study focused on the
characteristics of the two unstable Kelvin-Helmholtz modes supported by the base flow. As expected
at supersonic speeds, the first azimuthal Fourier modes ($m=1$) exhibited higher amplification than
their axisymmetric counterparts. The authors showed that, due to reduced velocity gradients, the
instability waves in the inner shear layer were less unstable than those in the outer shear
layer. The presence of the inner shear layer was found to affect the dynamics of the outer layer,
resulting in higher growth rates and lower phase velocities compared to equivalent single
jets. Consequently, the authors theoretically identified that supersonic IVP jets could emit
less mixing noise than the reference jet when the velocity ratio is small, the outer stream is
hotter than the inner stream, and the area ratio is small.\\
Given the similarities between the nozzle exhaust flow and IVP jets, it is reasonable to expect that
the dynamics of nozzle jet flow may also exhibit resemblances to the IVP jets one. However, it's worth
noting that in \cite{dahl1997noiseIII}, the study only considered cases where both flow streams were
supersonic, while in the nozzle exhaust, the inner flow is decelerated to subsonic speed through a
normal shock wave as depicted in figure \ref{fig_tic}. Additionally, recent studies have highlighted
the importance of guided jet modes (GJM) in the dynamics of supersonic jet flows. These waves were first mentioned by \cite{tam1989three}, only two years
before the study by \cite{dahl1997noiseIII}, and were therefore not discussed in the latter
paper. \vjt{Contrary the KH waves, these
  instability waves are neutral duct-like modes \citep{towne2017acoustic, schmidt2017wavepackets}
  that can have the interesting property of propagating in the upstream direction, towards the
  nozzle. This makes them perfect candidates to provide feedback in a resonance loop
  \citep{gojon2018oscillation, edgington2018upstream, jaunet2019dynamics, mancinelli2021complex,
    nogueira2022absolute}.}\\
To the best of the authors knowledge, there is no description of such waves in supersonic
over-expanded jet, and no model has yet been proposed to predict the occurrence of resonances
in the exhaust flow of convergent-divergent nozzle. Therefore, in this study, we propose
to investigate the linear dynamics of over-expanded nozzle exhaust flows. \vjt{A simplified
  framework of parallel flows is adopted and the study focuses on}
the effect of the base flow changes, mimicking variations in nozzle expansion regime, on the
characteristics of the most unstable waves and the guided jet modes (GJM), which are essential
components of some possible feedback resonance process.


\section{Dynamical Model}
\label{sec:models}
\subsection{Base flow modeling}
\subsubsection{Parametrization of the base flow}
\begin{figure}
  \centering
  \includegraphics[width=0.6\textwidth]{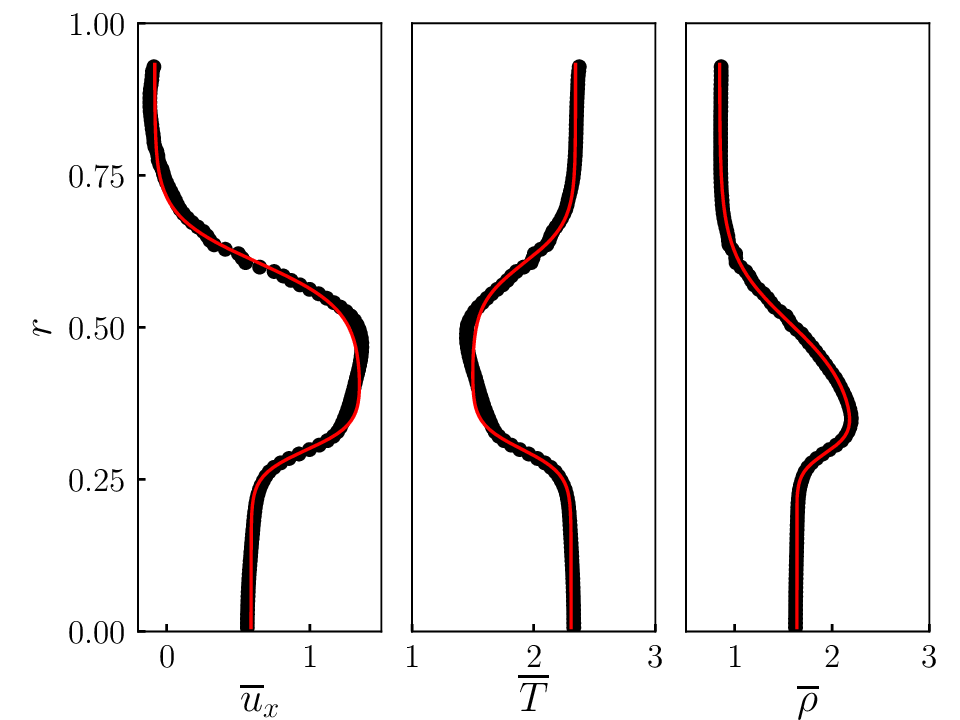}
  \caption{Typical base flow used for the study of local stability properties of the flow. The considered profiles are extracted at the distance from the throat $x/D = 1.6$ (where $D$ is the exit nozzle diameter), in between the two first Mach disks of the flow. Dots represent the numerical data and red lines are the fitted analytical functions.}
  \label{fig:base_flow}
\end{figure}

As can be seen in figure \ref{fig:base_flow}, the mean flow profiles downstream of the first Mach
disk, are, as expected, composed of two MLs: one separating the supersonic flow and the ambient one, the other one separating the subsonic core and the supersonic flow.
\vjt{A representative analytical formulation of this base flow is obtained by connecting three uniform
regions: the inner, the annular and the external one; with two hyperbolic tangent profiles to account for the mixing layers:}
\begin{eqnarray}
    q_{in}(r) & = & {q_i}\left[1 - \left(1 - \frac{q_a}{q_i}\right) \left(1+\tanh\left(\frac{2}{\theta_{i}}(r-R_i)\right)\right) \right] \nonumber \\
    q_{out}(r) & = & 1 - \frac{1}{2}\left(1-\frac{q_e}{q_a}\right)\left(1+\tanh\left(\frac{2}{\theta_{e}}(r-R_e)\right)\right) \nonumber\\
  q(r) & = & {q_{in}(r)} \cdot {q_{out}(r)},
             \label{eq:mean_model}
\end{eqnarray}
where $q$ here stands for $M$, $\bar{\rho}$ or $\bar{T}$. The indices i, a, and e respectively refer
to the inner, the annular and the external flow variable of interest. $(R_{i},
\theta_{qi})$,$(R_{e}, \theta_{qe})$ are the inner and external ML radial positions and thicknesses,
respectively. A regression of the analytical profiles onto numerical data available (DDES simulation previously done in
\cite{bakulu2021jet}) gives access to the parameters. The reader should note that different mixing
layer positions and thicknesses may be attributed to the mean velocity, density or temperature profile
when a direct fit to the numerical data is performed. As the flow velocity profile is concerned, this formulation is equivalent to the one
used in \cite{michalke1984survey}. This modelisation differs from the one chosen by
\cite{bhat1993effect} who preferred gaussian profiles instead of hyperbolic tangent ones.\\
Optimal analytical profiles are compared with the reference numerical
data in figure \ref{fig:base_flow}, demonstrating the ability of the analytical functions chosen to
represent the computed flow field. \vjt{The reader may note that the velocity and temperature
  profiles are not constant in the supersonic stream. This cannot be accounted for using the chosen
  analytical formulation and the modeled profiles show a plateau in this region. It is believed that
this simplification will not lead to drastic change in the results as the velocity and temperature
variation within the supersonic stream is very small compared to the variations in the mixing layers.}

\subsubsection{Obtaining the base flow parameters}
\label{sec:mean_flow_model}
\begin{figure}
  \centering
  \includegraphics[width=0.6\textwidth]{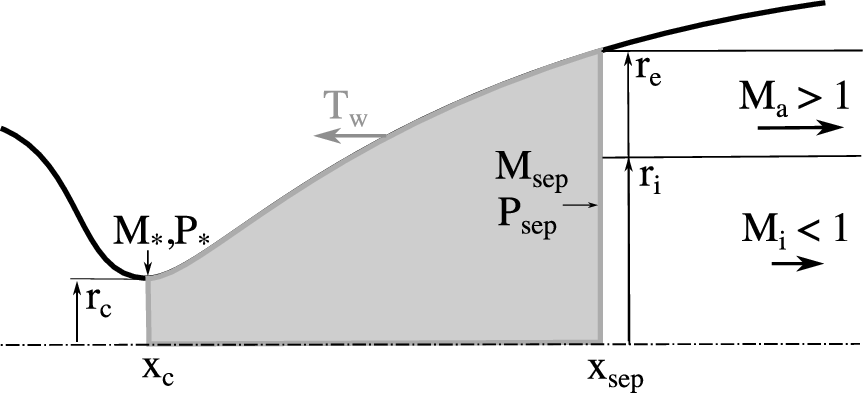}
  \caption{Schematic of the control volume (C.V.) employed to parameterize the flow field.}
  \label{fig:control-volume}
\end{figure}

Annular supersonic flows are commonly encountered when flows exit convergent-divergent nozzles and
often exhibit a sufficiently large Mach disk within their core \citep{hadjadj2009nozzle}. A
simplified  analytical description of the shock waves network  is proposed in various studies, using
mass and momentum conservation, for example in for example in \cite{chow1975mach} and
\cite{li1998mach}, we opted to derive an even simpler description of the flow, using quasi 1D relations, to obtain the
parameters that our finite
thickness model requires. Indeed, the earlier-presented base flows can be parameterized with the following set of parameters:
\emph{($r_i, r_e, M_i, M_e, M_a, T_i, T_a$)}. \vjt{In this section, we will often use comas to indicate
  interchangeable indices for sake conciseness. For example, the former vector of parameters could have been
  written {($r_{i,e}, M_{i,e,a}, T_{i,a}$)} with this notation.}
Basic fluid mechanics principles reveal that these
parameters are interconnected and we cannot assume arbitrary values. Hence, to attain a meaningful
parameter set, a simplified model of the flow inside the nozzle is built, assuming
quasi-one-dimensional behavior from the throat to the separation point, where the annular supersonic
flow is presumed to originate. The simplified flow configuration, and the associated control
volumes, that we propose to use in that purpose is illustrated in Figure \ref{fig:control-volume}.

By applying mass conservation to a control volume bounded by the nozzle throat and the attached flow (as depicted in gray in Figure \ref{fig:control-volume}), the following equation can be derived:
\begin{eqnarray}
  \rho_* U_* D_c^2 &=&  \rho_i U_i D_i^2 + \rho_a U_a (1-D_i^2),
\end{eqnarray}
where $(\cdot)_*$ denotes sonic variables at the throat, and $D_{i,c} = \frac{D_{i,c}}{D_e}$ represents
the normalized radius of the inner ($\cdot_i$) vortex sheet or the throat ($\cdot_c$) relative to
the external vortex sheet.\\
Noting $\mathcal{M}$, the local Mach number, differing from the acoustic Mach number employed in the linearized model: $\mathcal{M} = M
\frac{c_\infty}{c}$, where $c$ is the local speed of sound, the following relation can be easily
obtained from the previous equation:
\begin{eqnarray}
  {\left(1+\frac{\gamma-1}{2}\right)^{-\frac{\gamma+1}{2(\gamma-1)}}}
  \frac{P_{t0}}{\sqrt{T_{t0}}} D_c^2 &=& \frac{P_i}{\sqrt{T_i}} \mathcal{M}_i D_i^2 +
                                           \frac{P_a}{\sqrt{T_a}} \mathcal{M}_a (1-D_i^2),
\end{eqnarray}
\vjt{where $P_{t0}$ and $T_{t0}$ are the total pressure and temperature of the flow and $P_{i,a}$ and
$T_{i,a}$ are the static pressure and temperature of inner ($\cdot_,i$) and annular ($\cdot{,a}$) streams.}
Assuming an isobaric jet for simplification \textit{i.e.}, $P_a = P_i = P_\infty$, the mass
conservation equation can be further simplified (see the appendix \ref{appendix} for further details):
\begin{eqnarray}
\beta_m {D_c^2}   &=&  (\mu_i - \mu_a) {D_i^2}  +  \mu_a,
\end{eqnarray}
where:
\begin{eqnarray}
  \mu_{i,a}& =&  { \mathcal{M}_{i,a} \sqrt{1+\frac{\gamma-1}{2}\mathcal{M}_{i,a}^2}} \nonumber\\
  \beta_m  &=& {\left(1+\frac{\gamma-1}{2}\right)^{-\frac{\gamma+1}{2(\gamma-1)}}}
  \frac{P_{t0}}{P_\infty}.\nonumber
\end{eqnarray}

In essence, the mass conservation equation establishes a connection between the nozzle pressure
ratio $NPR = {P_{t0}}/{P_\infty}$, its geometry and the Mach numbers of both the annular and
internal jets. The NPR is dictated by the nozzle's operational regime, $D_c$ is given by nozzle
design considerations.

Considering momentum conservation along the axial direction within the same control volume, while neglecting body and viscous forces, leads to the following equation:
\begin{eqnarray}
  {\beta_q} D_c^2 - \frac{F_w}{P_\infty}
   & = & \gamma D_i^2 \left( \mathcal{M}_i^2 - \mathcal{M}_a^2 \right) + \left(\gamma \mathcal{M}_a^2 + 1\right),
\end{eqnarray}
where:
\begin{equation}
   \beta_q = (1+\gamma)\left(1+\frac{\gamma-1}{2}\right)^{\frac{-\gamma}{\gamma-1}}
   \frac{P_{t0}}{P_\infty},
\end{equation}
and $F_w$ represents the pressure forces acting on the nozzle wall. It's important to note that $F_w$ depends on the nozzle geometry and can be approximated using characteristic methods. By solving this equation, we can derive an analytical expression for $D_i^2$:
\begin{eqnarray}
    D_i^2 &=&\frac{1}{\gamma \left(\mathcal{M}_i^2 -
              \mathcal{M}_a^2\right)}\left[ {\beta_q} D_c^2
              - \frac{F_w}{P_\infty} - \left( \gamma \mathcal{M}_a^2 +1 \right)\right] \nonumber,
\end{eqnarray}
which can subsequently be substituted into the mass conservation equation. This substitution leads
to an equation for the annular Mach number, providing that $\mathcal{M}_{i}$ and $F_w$ are known:
\begin{eqnarray}
  {\frac{\mu_i - \mu_a}{\gamma \left(\mathcal{M}_i^2 - \mathcal{M}_a^2\right)}
  \left[ {\beta_q}{D_c^2} - \frac{F_w}{P_\infty}
  - \left( \gamma \mathcal{M}_a^2 + 1 \right)\right] }  +
  \mu_a - \beta_m    &=& 0,
                                 \label{eqn:annular_Mach}
\end{eqnarray}
where:
\begin{eqnarray}
  \mu_{i,a} & =&  { \mathcal{M}_{i,a} \sqrt{1+\frac{\gamma-1}{2}\mathcal{M}_{i,a}^2}}\nonumber\\
  \beta_m  & =& {\left(1+\frac{\gamma-1}{2}\right)^{-\frac{\gamma+1}{2(\gamma-1)}}}
  \frac{P_{t0}}{P_\infty} \nonumber\\
  \beta_q  & =& (1+\gamma)\left(1+\frac{\gamma-1}{2}\right)^{\frac{-\gamma}{\gamma-1}}
  \frac{P_{t0}}{P_\infty}. \nonumber
\end{eqnarray}

It is worth mentioning that once $M_{i,a}$ are known, $T_{i,a}$ can be determined via the usual isentropic relations, using the typically known total temperature of the flow.

\subsubsection{Solving the base flow model}
Equation \ref{eqn:annular_Mach} can be solved to determine the Mach number of the annular flow,
adhering to the fundamental principles of fluid mechanics. This can be achieved for any given input
parameter set containing the nozzle pressure ratio ($NPR = \frac{P_{t0}}{P_\infty}$), throat
diameter ($D_c$), inner subsonic Mach number ($\mathcal{M}_i$), and nozzle thrust ($F_w$). These
parameters are intrinsically interconnected and must be chosen judiciously. We elucidate below how
these parameters can be defined based on the nozzle's pressure and Mach number profiles, along with
the application of a classical separation criterion. Let us assume that flow separation occurs at an
axial location within the nozzle, denoted as $x_{sep}$, where an isentropic expansion of the flow
inside the nozzle has resulted in a local Mach number $\mathcal{M}_{sep}$ and pressure $P_{sep}$.

Primarily, it is evident from Figure \ref{fig:control-volume} that our reference length scale,
$D_e$, can be equated to the diameter of the nozzle at the separation point $D_w$:
\begin{equation}
  D_e = D_w(x_{sep}).
\end{equation}
This implies that the reference length scale for the annular jet flow will vary according to the NPR.

Secondly, experimental observations suggest that the separation pressure $P_{sep}$ is related to the separation Mach number and the external pressure $P_{\infty}$ through a suitably chosen separation criterion:
\begin{equation}
  \frac{P_{sep}}{P_\infty} = F(\mathcal{M}_{sep}),
\end{equation}
where $F$ represents the separation criterion formula, enabling the determination of $P_{sep}$. Subsequently, assuming an isentropic expansion from the tank to the separation point, the NPR at the current flow condition can be calculated:
\begin{equation}
  \frac{P_{t0}}{P_\infty} = {F(\mathcal{M}_{sep})}{\left(1+\frac{\gamma-1}{2} \mathcal{M}_{sep}\right)^{\frac{\gamma}{\gamma-1}}}.
\end{equation}

An approximation for the inner Mach number $\mathcal{M}_i$ can be derived by considering that the Mach disk is formed at the upstream Mach number $\mathcal{M}_{sep}$:
\begin{equation}
  \mathcal{M}_i = \frac{(\gamma-1)\mathcal{M}_{sep}^2+2}{2\gamma \mathcal{M}_{sep}^2-(\gamma-1)}.
\end{equation}
It is important to note that this approximation leads to an overestimation of $\mathcal{M}_i$ since,
in real flow conditions, the Mach disk always forms downstream of the separation point, hence at
higher Mach numbers.

Finally, the thrust of the control volume, $F_w$, can be determined by integrating the wall pressure along the nozzle from the throat to the separation point $x_{sep}$:
\begin{equation}
  F_w = \int_{x_c}^{x_{sep}} \left(P_\infty - P_w \right) (\vec{n} \cdot \vec{i} \,) \; dS.
\end{equation}

\begin{figure}
  \centering
  \includegraphics[width=0.45\textwidth]{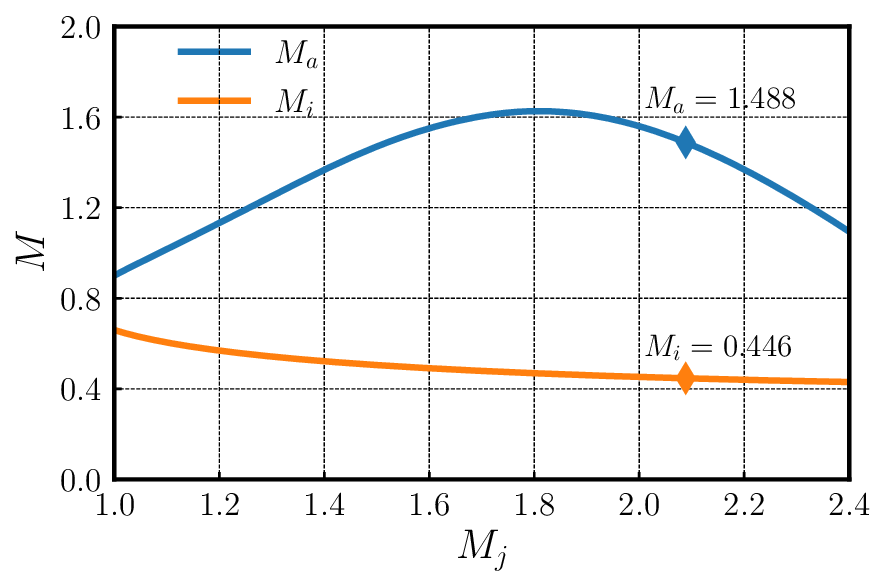}
  \includegraphics[width=0.45\textwidth]{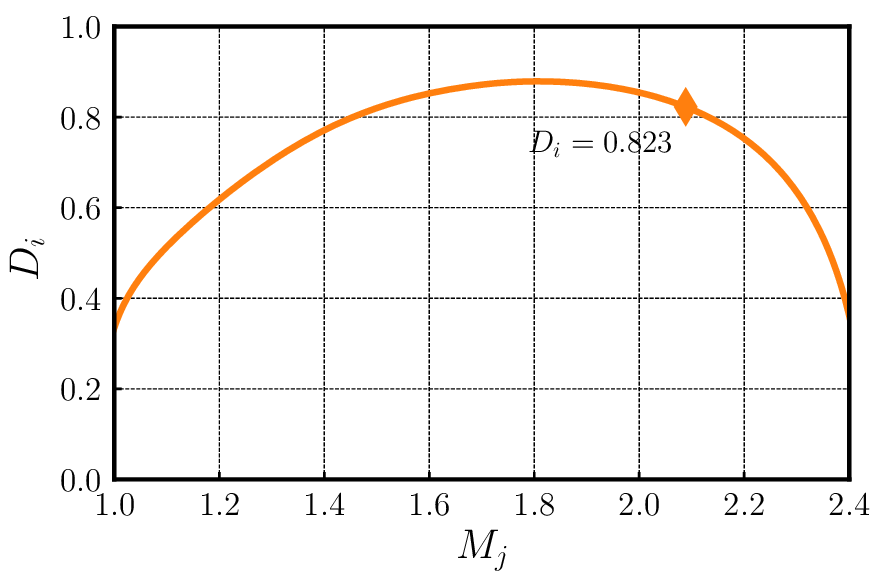}
  \caption{Variation of Mach numbers (left) and location of inner vortex sheet diameters (right) with the fully expanded jet Mach number.}
  \label{fig:model-Mach}
\end{figure}

The model is now complete and only requires the nozzle wall radius and Mach number profiles, which
can be computed using methods like the Method of Characteristics (MOC). An example using nozzle
characteristics from \cite{jaunet2017wall} is presented in the following section. The acoustic Mach
numbers of the vortex sheets and the inner radius satisfying the proposed model are depicted in
Figure \ref{fig:model-Mach} against the fully expanded Mach number of the flow $M_j$. $M_j$ can be
easily computed from the nozzle NPR using isentropic relations. The separation criterion from
\citet{Stark2005}, with slight modifications to better align with the experimental observations of
\cite{jaunet2017wall}) for this nozzle geometry, is employed. The decrease in $\mathcal{M}_i$ as the normal shock wave strengthens with downstream displacement of the separation, is in accordance with expectations. Conversely, neither $M_a$ nor $D_i$ display a monotonic behavior. This can be attributed to the nozzle wall profile approaching a near-horizontal orientation near the lip, resulting in saturation of thrust $F_w$. Another noteworthy behavior is the sudden decrease in $D_i$ as the separation approaches the nozzle lip, consistent with the behavior of overexpanded jets from TIC nozzles—akin to an approaching adaptation where the Mach disk diminishes or disappears.

Mach number and inner shear layer location values obtained for $M_j = 2.09$ are indicated as diamonds in Figure \ref{fig:model-Mach}. Table \ref{tab:modelVsDDES} compares the model's predictions with those observed in numerical simulations from \cite{bakulu2021jet}, specifically after the first Mach disk. The results are summarized in Table \ref{tab:modelVsDDES}. As seen, the proposed model accurately predicts Mach numbers within a $\pm$ 10\% range but tends to overestimate $D_i$. This discrepancy is due to assuming a horizontal flow exit from separation, which is erroneous as the flow is deflected through the separation shock wave. However, this representation acknowledges its simplified nature. Adjusting the empirical separation model, or using a more representative value for $D_e$, would yield slightly different values for $M_{i,a}$ and $D_i$. Despite these uncertainties, it is believed that the model remains valuable for understanding the overall dynamics of the flow.

\begin{table}
  \centering
    \begin{tabular}{cccc}
         & $M_i$ & $M_a$ & $D_i$\\
  DDES   & 0.40  & 1.42  & 0.48\\
  Model  & 0.446 & 1.488 & 0.82
    \end{tabular}
  \caption{Comparison of Mach numbers and inner shear layer locations obtained from the DDES of \cite{bakulu2021jet} and the proposed simplified model.}
  \label{tab:modelVsDDES}
\end{table}

In conclusion, we have formulated a comprehensive model that derives essential parameters for
feeding a simplified linear dynamical model of jet flows exiting from over-expanded nozzles. \vjt{This
model requires the \textit{a priori} knowledge of the static pressure and Mach number profiles along the nozzle and a
separation criterion. From this, all the necessary quantities to build a simplified mean annular
supersonic mean flow for various nozzle regimes can be retrieved}.


\subsection{Finite thickness dynamical model}

We propose to use the same approach as \citet{mancinelli2020complexvalued}, modelling the jet using locally-parallel linear stability theory. All variables are normalised by the fully expanded jet diameter $D_j$, the ambient density and speed of sound $\rho_\infty$ and $c_\infty$, respectively. The Reynolds decomposition,
\begin{equation}
    q'(x, r, \theta, t) = \overline{q}(r) + q(x, r, \theta, t),
\end{equation}
is applied to the flow-state vector $q'$, where the mean and fluctuating components are $\overline{q}$ and $q$, respectively. We assume the normal-mode ansatz,
\begin{equation}
q(x, r, \theta, t) = \hat{q}(r) e^{i(k x + m\theta-\omega t)}
\end{equation}
where $k$ is the streamwise wavenumber and $m$ is the azimuthal Fourier wavenumber. The
non-dimensional pulsation $\omega =  2 \pi St M_a$ is computed via the acoustic Mach number of the
jet $M_a = U_j/c_\infty$ and the Strouhal number of the flow $St = f{D_j}/{U_j}$, where $U_j$ and
$D_j$ are the equivalent fully expanded velocity of the flow and diameter of the jet, respectively.

Linearizing the Euler equation around the base flow, we obtain the compressible Rayleigh equation for pressure,
\begin{equation}
  \dddr{\hat{p}}
  + \left( \frac{1}{r} - \frac{2k}{\overline{u}_x k -\omega} \ddr{\overline{u}_x}
    -\frac{\gamma -1}{\gamma \overline{\rho}}\ddr{\overline{\rho}}
    + \frac{1}{\gamma \overline{T}}\ddr{\overline{T}} \right) \ddr{\hat{p}}
  -\left( k^2 + \frac{m^2}{r^2} - \frac{(\overline{u}_x k
        - \omega)^2}{(\gamma - 1) \overline{T}} \right) \hat{p} = 0,
  \label{rayleigh-eqn}
\end{equation}
where $\gamma$ is the specific heat ratio for a perfect gas. The solution of the linear stability
problem is obtained specifying a real or complex frequency $\omega$ and solving the resulting
augmented eigenvalue problem $k = k (\omega)$, with $\hat{p} (r)$ the associated pressure
eigenfunction. The eigenvalue problem is solved numerically by discretizing \ref{rayleigh-eqn}
in the radial direction using Chebyshev polynomials. A mapping function is used to non-uniformly
distribute the grid points such that they are dense in the region of shear
\citep{trefethen2000spectral}.\\
Although the model described above supports non-isobaric regime \citep{mancinelli2023reflection}, only
isobaric jets will be studied in the following. Moreover, only the first non-axisymmetric
fluctuating mode ($m=1$) is considered, for this azimuthal mode is the only one responsible
for side-loads in overexpanded nozzle flows \citep{Dumnov1996}, and for resonances observed in
experiments \citep{jaunet2017wall}.\\
\vjt{ The flow dynamics is also studied without taking the nozzle walls into account. This
  corresponds to a free jet configuration and we are looking for pressure waves
  vanishing at infinity. This choice is motivated by the fact that, in our previous
  observations, the downstream- and uptream-propagating waves signatures were observed as being dominant far downstream of
 the nozzle, where the influence of the nozzle wall is negligible.}

\subsection{Convergence of the linear stability calculation}
\begin{figure}
  \centering
  \includegraphics[trim={0cm 0cm 0cm 0cm},clip,
  width=0.49\textwidth]{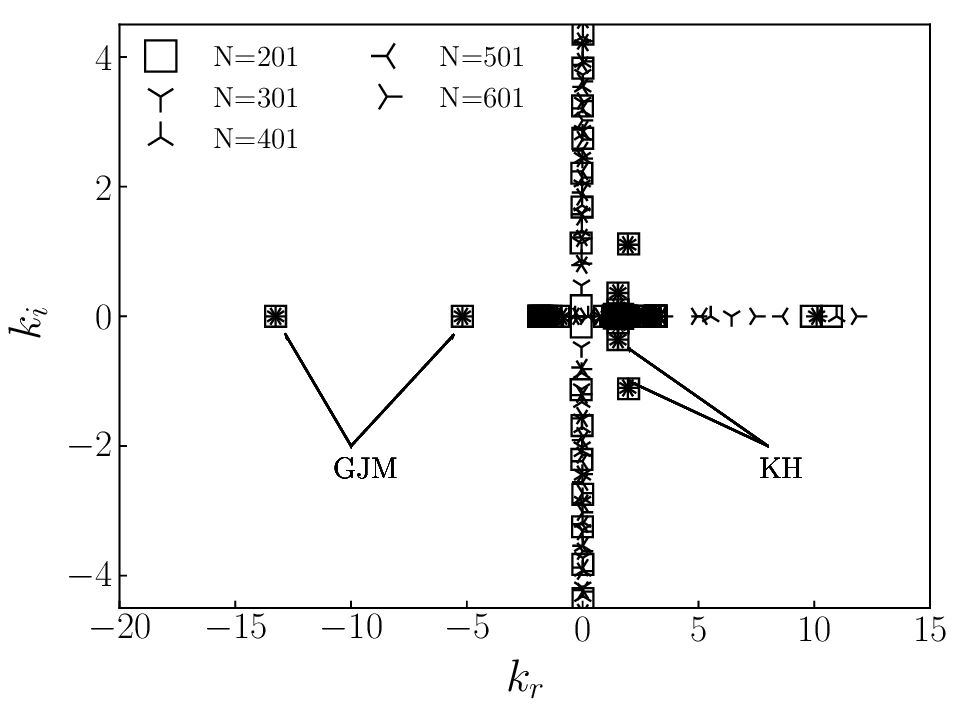}
  \includegraphics[trim={0cm 0cm 0cm 0cm},clip,
  width=0.49\textwidth]{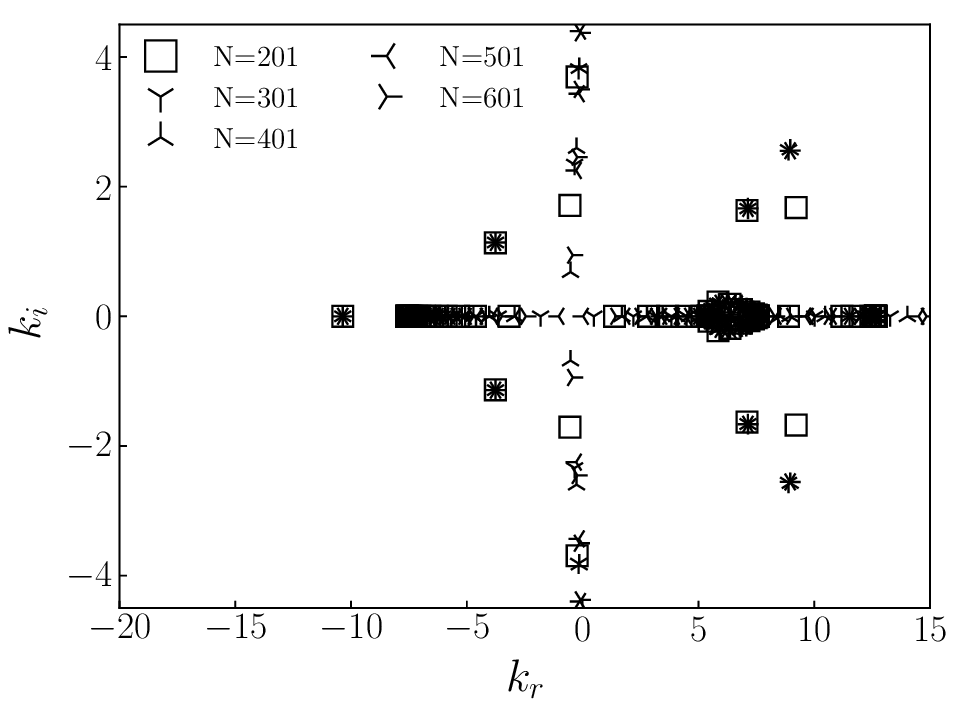}

  \caption{Convergence of the eigenvalue spectrum with respect to the number of Chebychev
    collocation points for $St = 0.1$ (left) and $St = 0.4$ (right).}
  \label{fig:conv}
\end{figure}

We present in figure \ref{fig:conv} the convergence of the eigenvalue spectrum with respect to the
number of Chebychev collocation points $N$ and the Strouhal number. \vjt{We indicated in figure
\ref{fig:conv} the position of the waves of interesrest in this paper, namely the guided jet modes
(GJM) and the two unstable waves ($k_i <0$) that lie in the $k_r>0$ portion of the spectrum and are
downstream travelling. These modes will be named inner and outer Kelvin-Helholtz (KH) waves in the
document for they have spatial support with a maximum located on both the inner and the outer mixing
layers, as expected for Kelvin-Helmholtz (KH) waves. This denomination seems logical although it
will be shown later that their spatial supports shows some differences, especially regarding their
decay in the radial direction. This might need a separate discussion by itself, but it is out of the
scope of the curent study.\\
The GJM, on the other hand, lie in the $k_r<0$ region and on the real axis, they
hence are neutral wave with negative
phase speed. With varying frequency, the position of these modes changes. The GJM moves along the
real axis and might separate from this axis and become evanescent. That is what wan be seen in
figure \ref{fig:conv} right, where two GJM can be seen near by the $(-5,\pm1.5)$ position. For a more
precise discussion on these modes, we refer the reader to \cite{towne2017acoustic}.}\\
The overlay of the different symbols in figure \ref{fig:conv} shows that the convergence is more
difficult to reach at high frequencies, especially for the
Kelvin-Helmholtz modes, but the spectrum seems to have converged above $N=301$ points for the
frequencies of interest in this paper. Therefore, a total number of $N=401$ collocation points is
used for all the computations in this study.

\section{Dynamics of annular supersonic jets}
\label{sec:dynamics}
Contrary to a single stream jet flow, the base flow of a annular jet requires more parameters to be perfectly described. The characteristic features of the base flow (mixing layer thickness or position) may evolve with the nozzle pressure ratio as well and, therefore, their relative influence on the stability spectrum needs to be evaluated before one may explore the conditions for resonance. For this purpose, we use the analytical profile described above and explore the influence of the main parameters on the stability properties of the jet.

\subsection{Effect of radial position}
\begin{figure}
  \centering
  \includegraphics[trim={0cm 0cm 0cm 0cm},clip, width=0.7\textwidth]{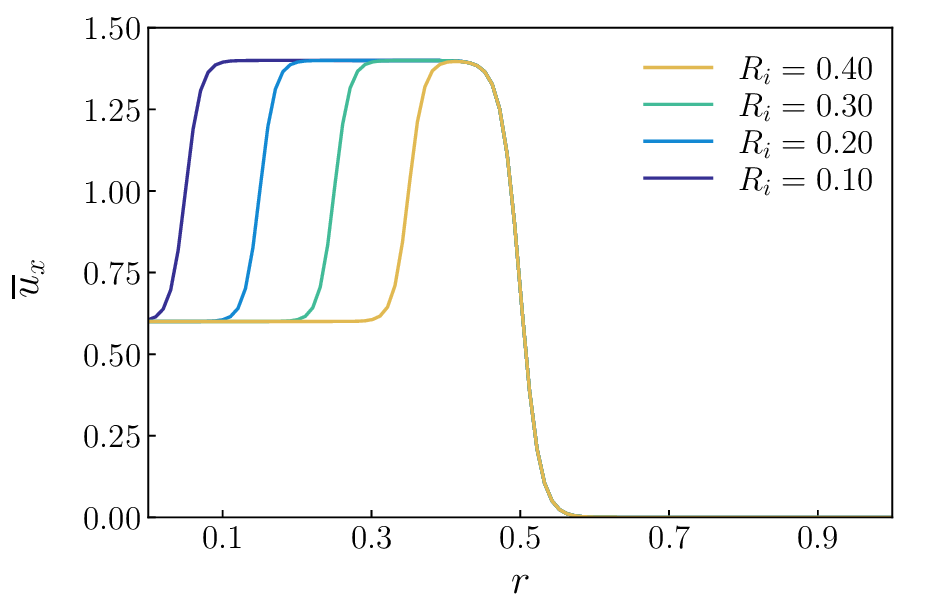}
  \caption{Base flows used to study the effect of the position of the inner mixing layer on the flow dynamics.}
  \label{fig:baseflow_varRi}
\end{figure}

The influence of the relative positions of the mixing layers is first studied by assuming a
constant reference velocity levels in the subsonic core and the supersonic stream. We perform linear
stability analyses varying the position of the inner mixing layer and keeping the one of the outer
mixing layer constant, as presented in Figure
\ref{fig:baseflow_varRi}. The inner, annular, and external Mach numbers are chosen to approximately
match the numerical simulations downstream of the first Mach disk: $M_i = 0.6$, $M_a = 1.4$, and
$M_e = 0$. The density profile was computed using the Crocco-Busemann relation
\citep{michalke1984survey}. A total of 36 base flows is used to finely study the changes in the
corresponding linear dynamics.

\begin{figure}
  \centering
  \includegraphics[trim={0cm 0cm 0cm 0cm},clip,width=0.47\textwidth]
  {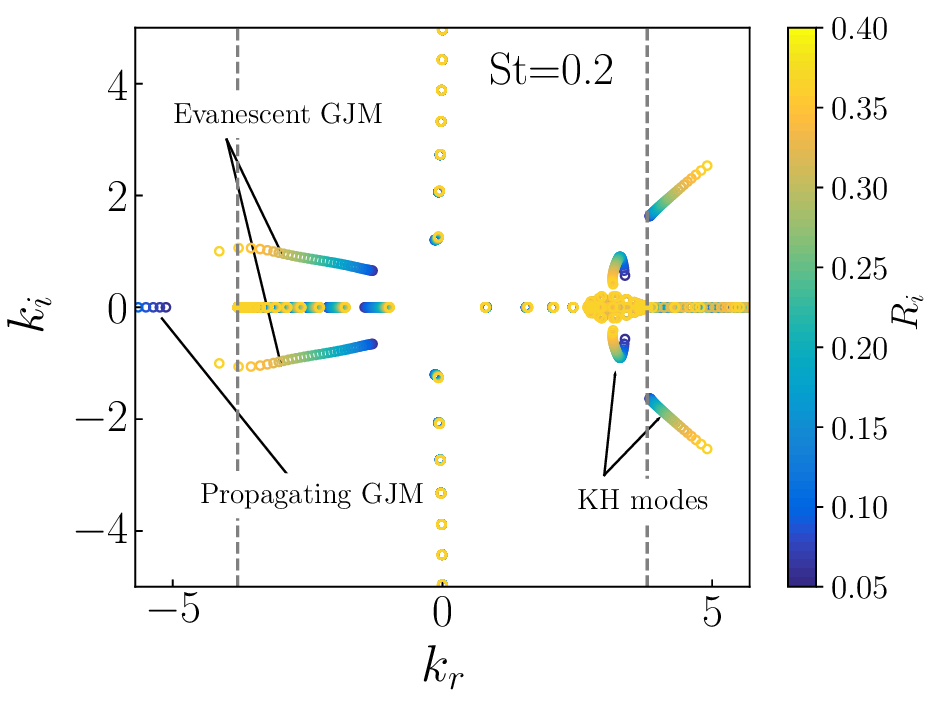}
  \includegraphics[trim={0cm 0cm 0cm 0cm},clip, width=0.47\textwidth]
  {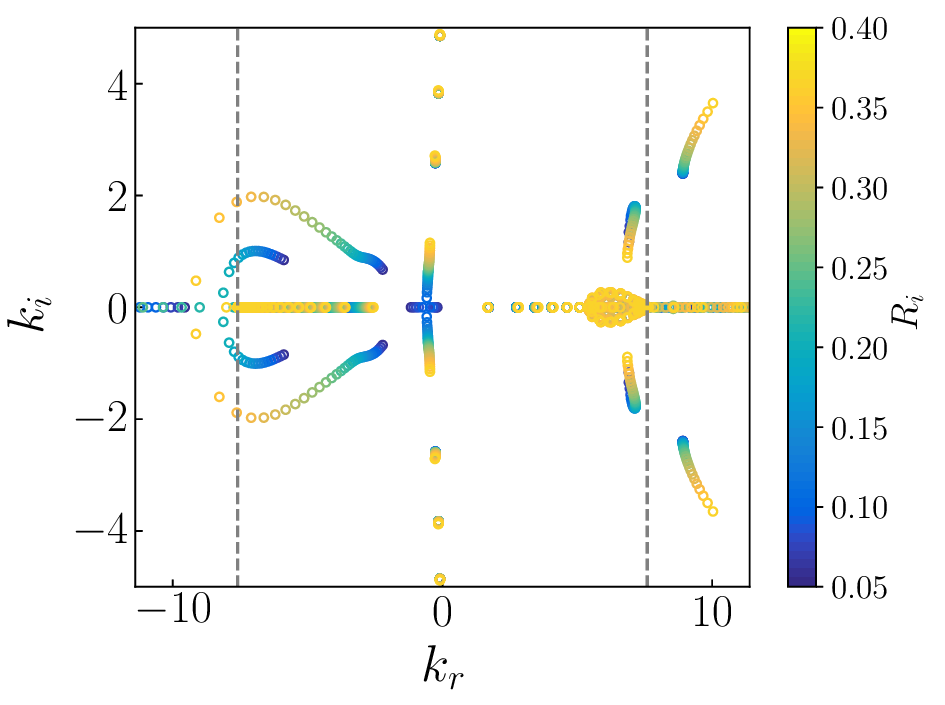}
  \caption{Eigenspectra computed at $St=0.2$ and $St=0.4$ for various $R_i$. The dashed lines indicate the wavenumber corresponding to sonic phase speed.}
  \label{fig:eigenspectrum}
\end{figure}

Typical eigenspectra obtained at $St=0.2$ and $m=1$ are presented in Figure
\ref{fig:eigenspectrum}. \vjt{On the right-hand side of the figure ($k_r > 0$), two groups of unstable
eigenvalues (\textit{i.e.} belonging to the $k_i < 0$ half plane) can be observed. They
correspond to the Kelvin-Helmholtz (K-H) modes of both the inner and outer mixing layers, as was
found by \cite{dahl1997noiseIII} for supersonic IVP jets.} The inner K-H mode is the least unstable
of the two. The vertical ($k_r = 0$) and horizontal ($k_i=0$) parts of the continuous acoustic
branch are clearly visible, with a relatively weak influence of $R_i$ on the loci of the
corresponding eigenvalues. Propagating and evanescent guided jet modes \citep{towne2017acoustic} can also
be identified in the $k_r < 0$ half plane. Thus, the annular supersonic jets appear to exhibit
similar features to those of a classical top-hat supersonic jet. In the following, we will focus on
extracting the effect of the inner mixing layer position on the characteristics of these waves.

\subsubsection{Kelvin-Helmholtz mode}
\begin{figure}
  \centering
  \includegraphics[trim={0cm 0cm 0cm 0cm},clip, width=0.6\textwidth]{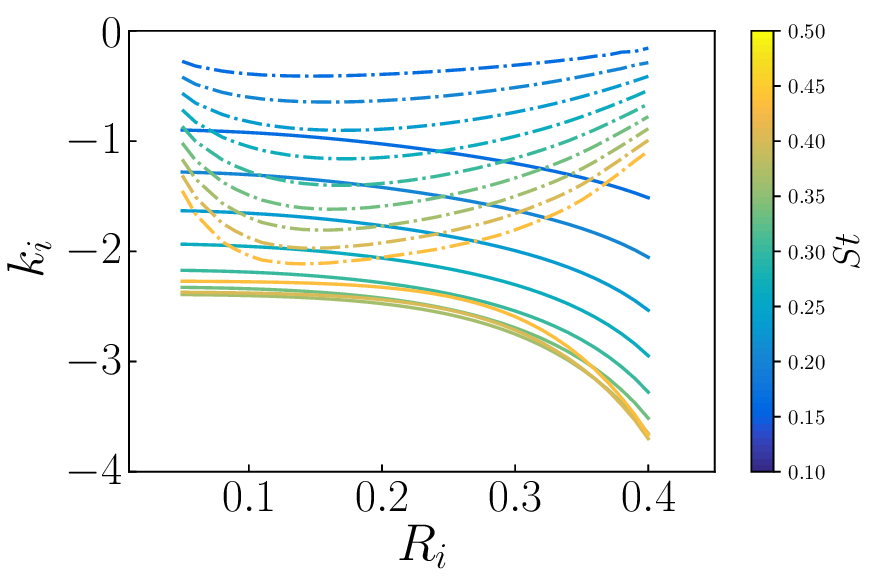}
  \caption{Growth rate of the Kelvin-Helmholtz mode as a function of the Strouhal number and the inner mixing layer location $R_i$. Dashed lines and solid lines correspond to the inner and outer K-H modes, respectively.}
  \label{fig:kh_growth}
\end{figure}

In Figure \ref{fig:kh_growth}, we plot the growth rate evolution of both K-H modes as a function of
$R_i$ for various Strouhal numbers. The outer K-H mode is always more unstable than the inner one
for all the configurations investigated in this study. This is expected considering that the
velocity gradient is stronger for the outer mixing layer than for the inner one. The outer mode
becomes even more unstable as $R_i$ increases, although at high frequencies its growth rate shows a
plateau at low values of $R_i$. On the other hand, the inner mixing layer instability shows an
optimal growth rate at specific $R_i$ values. This trend suggests that there may exist specific base
flows and frequencies for which the inner and outer K-H modes possess equivalent growth rates. In
some cases, the inner mode can even become more unstable than the outer one. However, this is more
likely to occur at higher frequencies and might not be relevant for the initial nozzle flow problem,
where resonance at lower frequencies is observed.

\begin{figure}
  \centering
  \includegraphics[trim={0cm 0cm 0cm 0cm},clip, width=0.49\textwidth]{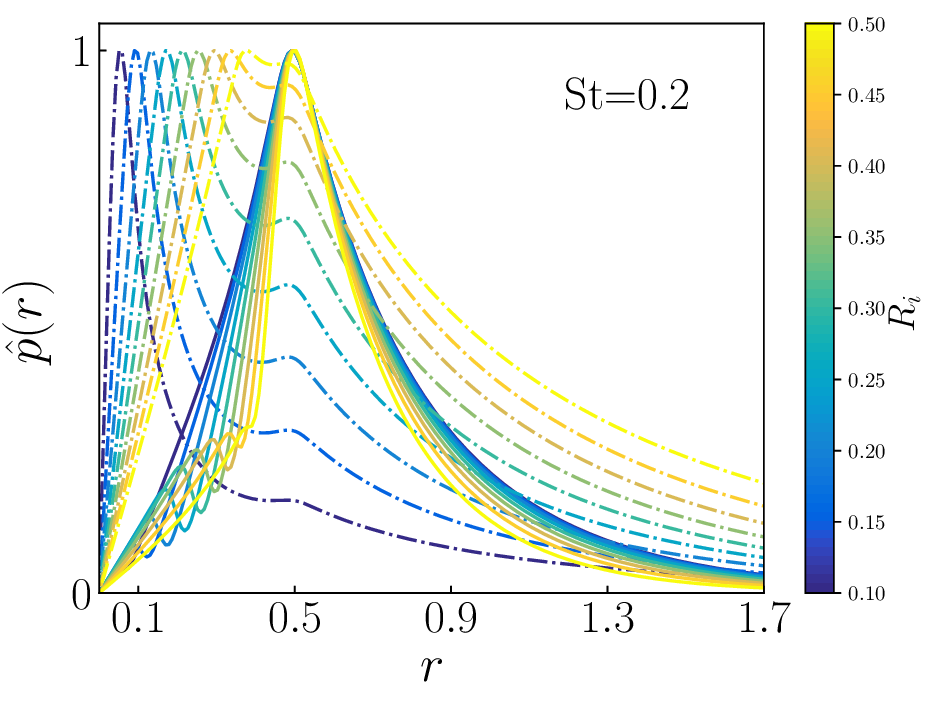}
  \includegraphics[trim={0cm 0cm 0cm 0cm},clip, width=0.49\textwidth]{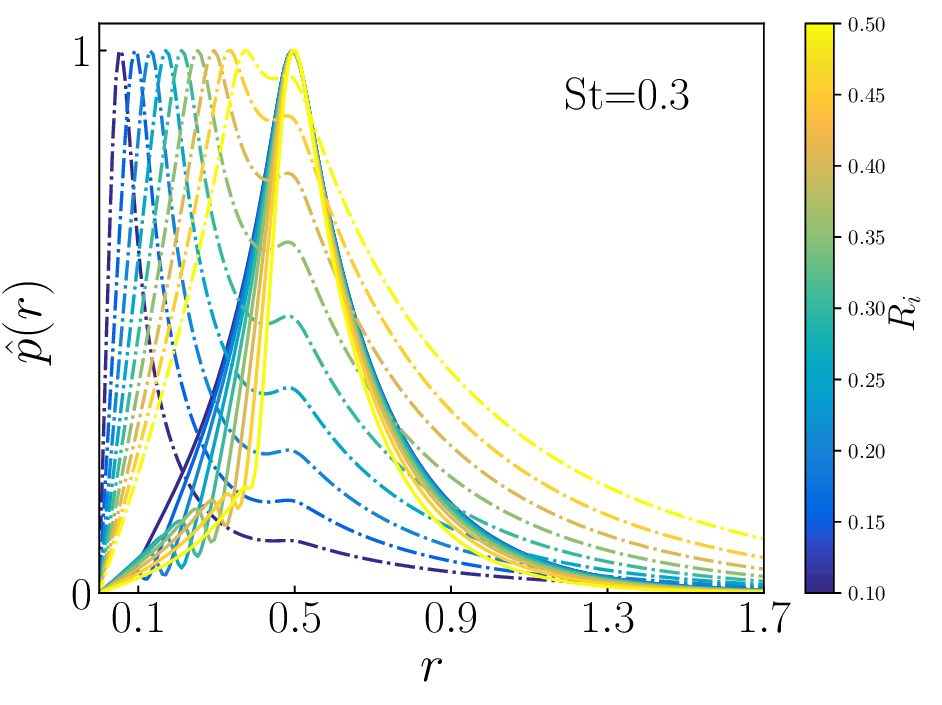}
  \caption{Eigenfunctions associated with the inner (dashed) and outer (solid) K-H modes, normalized with respect to their maximum values, for $St=0.2$ and $St=0.4$.}
  \label{fig:kh_func}
\end{figure}

The pressure eigenfunctions associated with the two identified K-H modes are presented in Figure
\ref{fig:kh_func} for $St=0.2$. As expected, the eigenfunction is maximum at the mixing layer
location and is zero at the centerline, as we focus on the $m=1$ anti-symmetric wavenumber. The
outer K-H mode does not seem to be significantly affected by the inner mixing layer, unlike the
inner mode whose support spreads further to the outside the outer mixing layer as the two mixing layers get closer. Interestingly, the radial support of the inner K-H wave decays less rapidly than that of the outer one. This allows the inner mode to be detected quite far from the jet and enables it to exchange energy with feedback waves outside of the supersonic annular region. This aligns well with the observations of \citet{bakulu2021jet}, who found that the inner mixing layer supports most of the perturbations associated with the resonance process in their flow.

\subsubsection{Guided jet modes}
\begin{figure}
  \centering
  \includegraphics[trim={0cm 0cm 0cm 0cm},clip,
  width=0.4\textwidth]{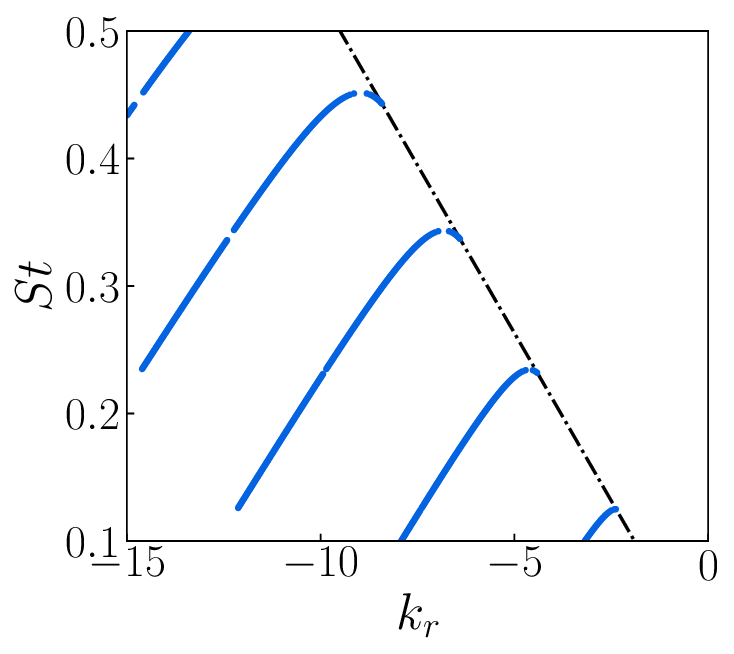}
  \includegraphics[trim={0cm 0cm 0cm 0cm},clip,
  width=0.4\textwidth]{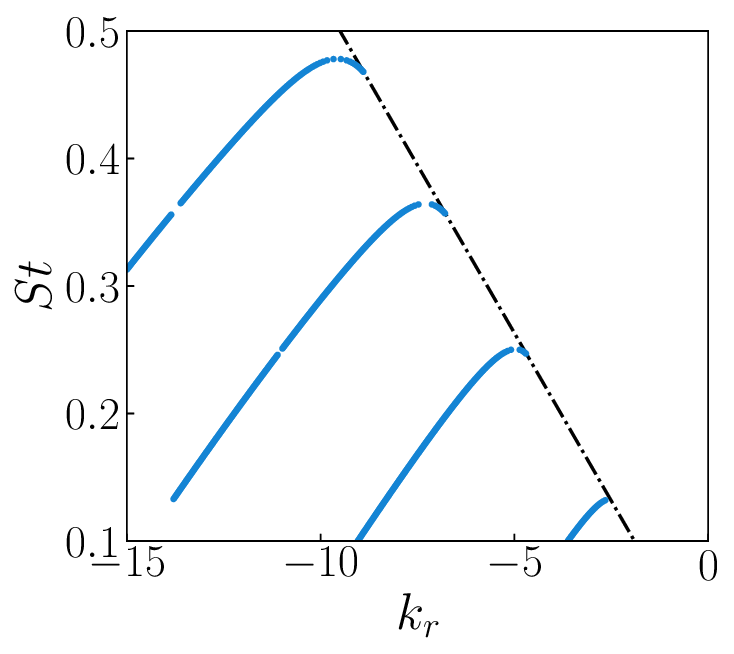}\\
  \includegraphics[trim={0cm 0cm 0cm 0cm},clip,
  width=0.4\textwidth]{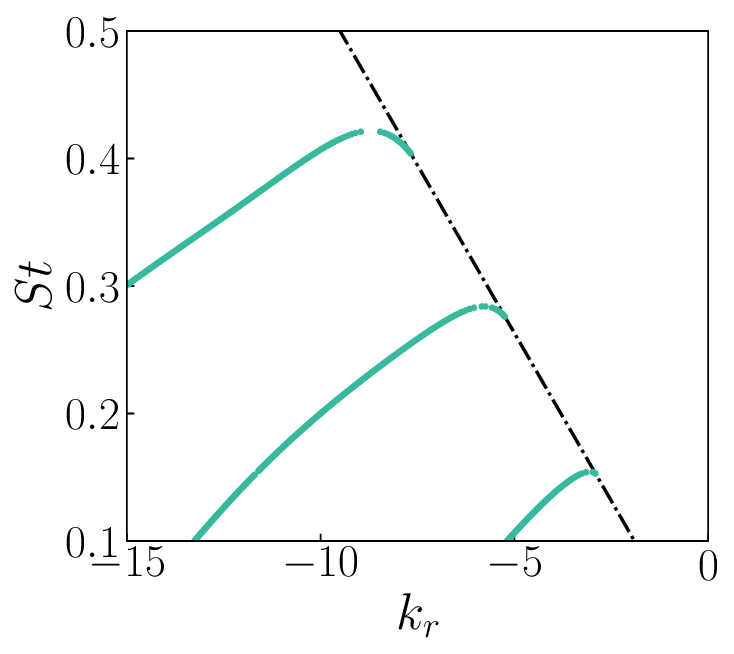}
  \includegraphics[trim={0cm 0cm 0cm 0cm},clip,
  width=0.4\textwidth]{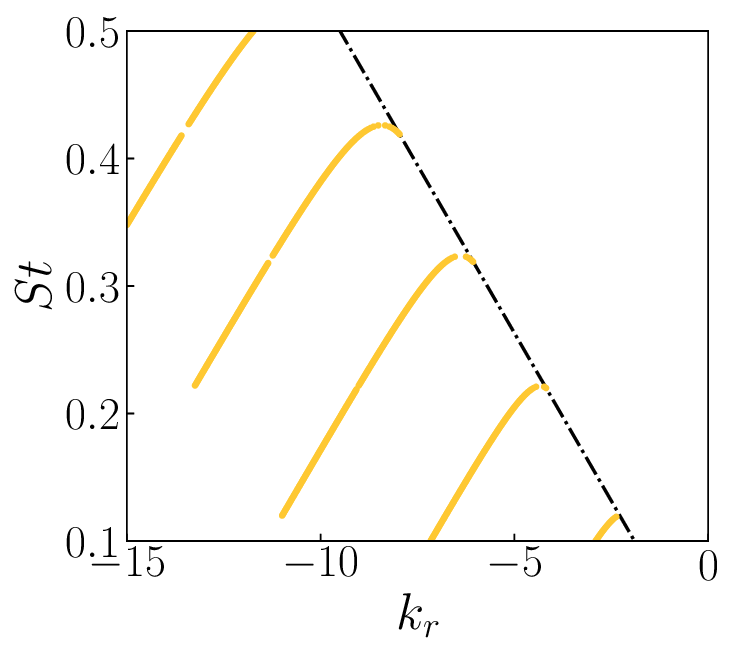}
  \caption{Neutral mode dispersion relation for various positions of the inner mixing layer: $R_i =
    0.05$ (top left),$R_i = 0.1$ (top right),$R_i = 0.2$ (bottom left),$R_i = 0.35$ (bottom
    right). The dash-dotted line correspond to the acoustic waves dispersion relation.}
  \label{fig:disprel_punk}
\end{figure}

Guided jet modes have been shown to play a significant role in the resonance of screeching and
impinging jets (see, for example, \cite{gojon2018oscillation, jaunet2019dynamics,
  mancinelli2021complex, edgington2018upstream}). Their dispersion relation is a key characteristic
that explains cut-on and cut-off frequencies, the transition from axisymmetric to helical modes, and
improves the model predictions \citep{mancinelli2019screech, mancinelli2021complex,
  nogueira2022absolute}. \vjt{As explained previously, their eigenvalues lie on the real axis of the
  spectrum until they become evanescent. Figure \ref{fig:eigenspectrum} shows that the annular jet configuration, as encountered
in the exhaust of over-expanded CD nozzles, also supports such type of waves.
  In the $(k_r,St)$ plane, this occurs at the summit of a given
branch. At supersonic speeds, jets support both upstream propagating, denoted $k_p^-$, and
downstream propagating, denoted $k_p^+$, GJM. This distinction is made based on the sign of their group
velocity $u_g = \frac{\partial \omega}{\partial k}$, hence corresponding to the local slope of their
dispersion relation presented in figure \ref{fig:disprel_punk}. The $k_p^-$ can be seen close
to the acoustic waves dispersion relation and is the mode that
can carry energy in the upstream direction.} The
dispersion relation of these waves is plotted in Figure \ref{fig:disprel_punk} in the $St-k_r$ plane
for various positions of the inner mixing layer.

As expected, at each $R_i$, the neutral waves form families of modes hierarchically ordered by their radial supports: higher frequencies correspond to higher radial orders (i.e., the number of nodes and antinodes, see \cite{tam1989three}). As seen, the position of the inner mixing layer has a strong impact on the neutral modes' dispersion relation: as $R_i$ increases, the frequencies at which the modes are encountered also increase. Furthermore, the group velocity of the $k_p^+$ mode decreases with increasing $R_i$. This can be understood by recognizing that for $R_i=0$ or $R_i=R_e$, the base flow is close to a supersonic top-hat jet or a subsonic one, respectively. As can be seen, both group and phase velocities of the guided jet modes decrease with increasing $R_i$. This is expected from the decrease of the overall average speed of the jet when $R_i$ increases. The neutral modes' behavior seems to lie between what is expected for a subsonic $M=M_i<1.0$ jet and a supersonic $M=M_a>1.0$ one.\\
Moreover, we observe that $R_i$ impacts the domain of existence of the upstream propagating neutral modes, the modes in the dispersion relation with negative group velocity. First, the domain where $d\omega/dk<0$ shifts towards higher Strouhal numbers with increasing $R_i$. Second, the range of Strouhal numbers where these neutral modes are encountered varies significantly with $R_i$, larger $R_i$ providing a wider range of existence, therefore offering more solutions for possible resonances.

\begin{figure}
  \centering
  \includegraphics[trim={0cm 0cm 0cm 0cm},clip, width=0.47\textwidth]{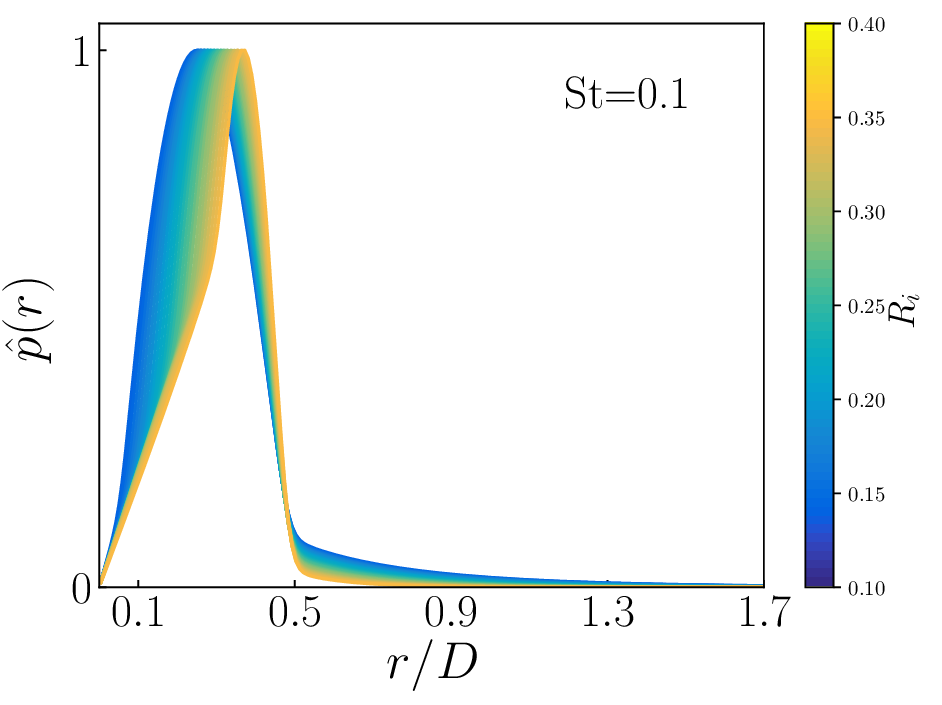}
  \includegraphics[trim={0cm 0cm 0cm 0cm},clip, width=0.47\textwidth]{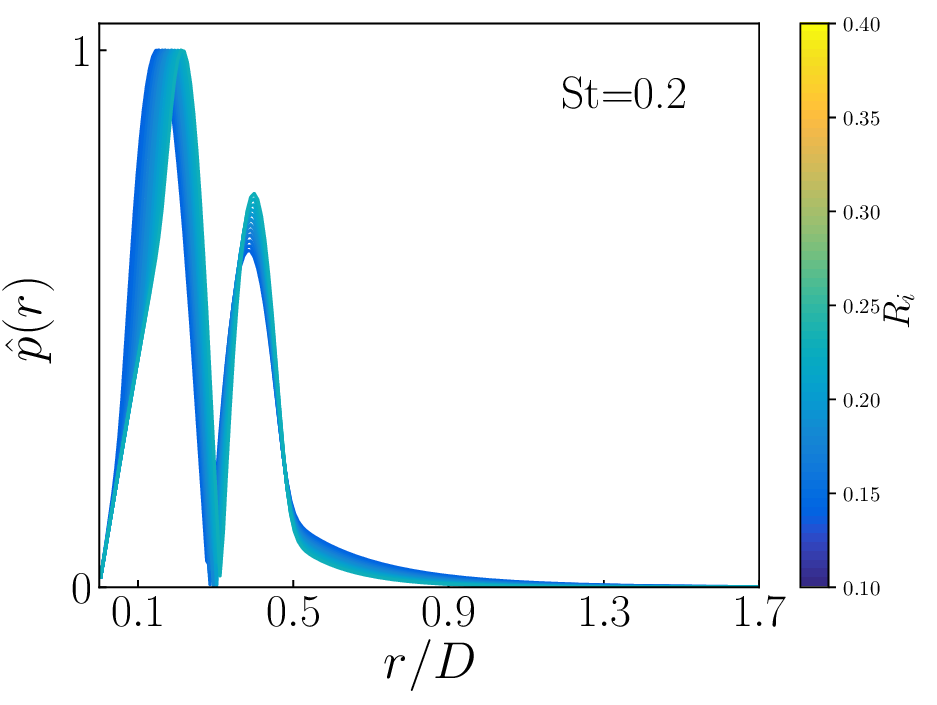}
  \caption{Eigenfunctions associated with the guided jet modes, normalized with respect to their maximum values, for $St=0.1$ and $St=0.2$.}
  \label{fig:kp_func}
\end{figure}

The eigenfunctions associated with the upstream propagating neutral modes are presented in Figure
\ref{fig:kp_func} for $St=0.1$ and $St=0.2$. As expected, the radial supports of these modes share
common features with Bessel functions. At low Strouhal numbers, the eigenfunctions show one
antinode, while for higher frequencies, more antinodes can be observed. Surprisingly, very little
change in the radial support is observed for varying $R_i$. This is due to the fact that these modes
are duct-like modes, so that their support is mostly independent of the base flow, in contrast to their wavenumber. It is important to notice that they also have support outside the jet, similar to the inner K-H wave. This allows these waves to interact and for the upstream mode to carry energy upstream, potentially closing a feedback loop.

\subsection{Effect of mixing layer thickness}
\begin{figure}
  \centering
  \includegraphics[trim={0cm 0cm 0cm 0cm},clip, width=0.7\textwidth]{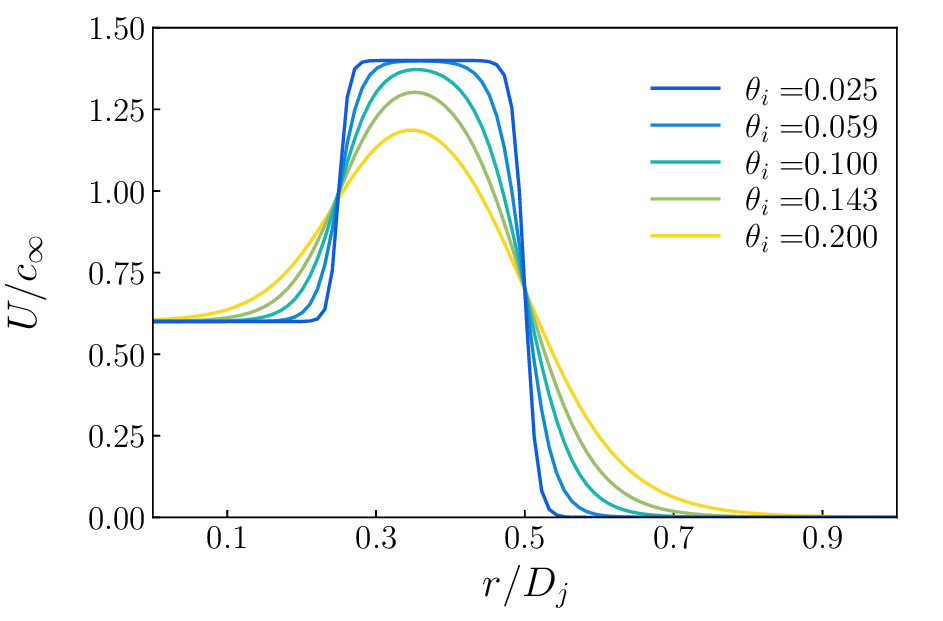}
  \caption{Base flows used to study the effect of mixing layer thickness.}
  \label{fig:baseflow_varT}
\end{figure}

In order to explore the influence of the thicknesses of the mixing layers on the stability
properties of the flow, we vary the base flow as presented in Figure \ref{fig:baseflow_varT}. Both
mixing layers are centered around $D_j/2$ and $D_j/4$. The jet bulk velocities are chosen to be
constant and comparable to those observed in the nozzle after the first Mach disk for the case
previously studied at $M_j=2.1$.

\subsubsection{Kelvin-Helmholtz mode}
\begin{figure}
  \centering
  \includegraphics[trim={0cm 0cm 0cm 0cm},clip, width=0.7\textwidth]{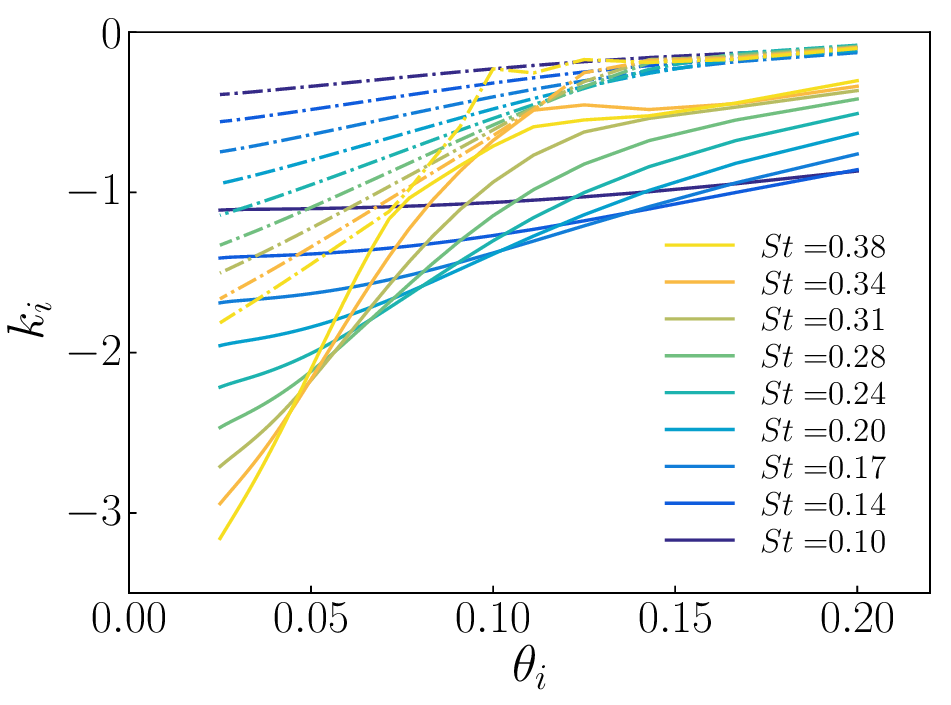}
  \caption{Growth rates of both the inner (\dashed) and outer (\full) Kelvin-Helmholtz modes as functions of the mixing layer thicknesses and the Strouhal number.}
  \label{fig:KH_growth_varT}
\end{figure}

Figure \ref{fig:KH_growth_varT} shows the growth rate of both KH modes with varying thicknesses and
Strouhal numbers. As expected, larger mixing layer thicknesses result in lower growth rates of the
KH mode. For the base flows and frequencies considered here, the inner mixing layer is more stable
than the outer one. This is due to their difference in velocity gradient. The outer mixing layer
separates a supersonic stream from a quiescent atmosphere, while the inner one separates two rapid
flows. Interestingly, for relatively high frequencies (above $St=0.2$), increasing the thickness
stabilizes the outer mixing layer more quickly than the inner one. Thus, there might be flow
configurations for wich both KH modes exhibit comparable growth rates. Depending on the frequency and
base flow, the dynamics may not be entirely dominated by the outer KH mode, as one might initially
assume.

\begin{figure}
  \centering
  \includegraphics[trim={0cm 0cm 0cm 0cm},clip, width=0.47\textwidth]{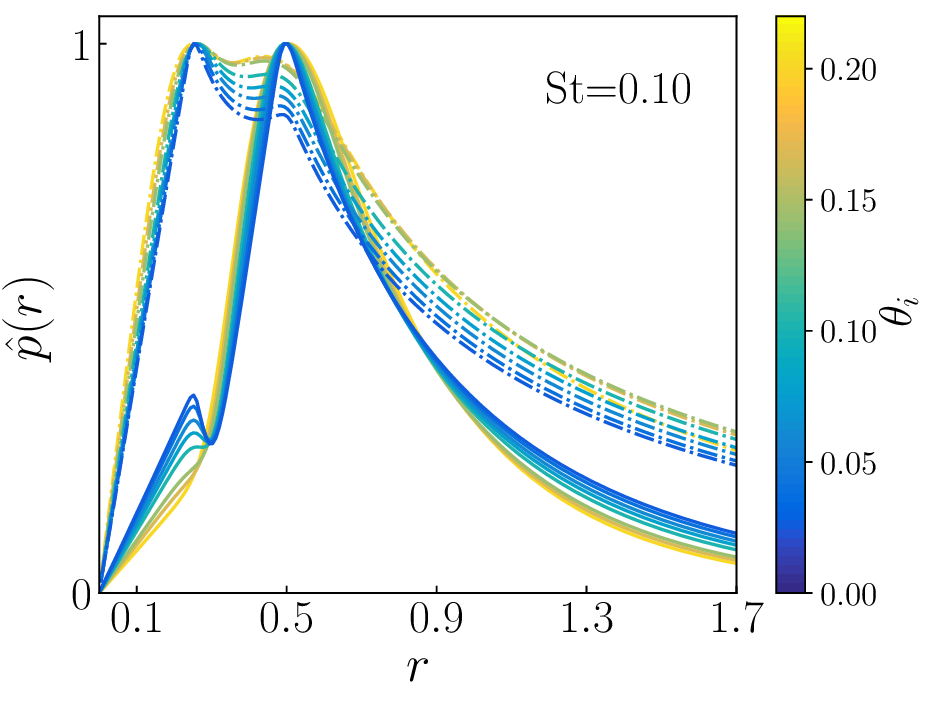}
  \includegraphics[trim={0cm 0cm 0cm 0cm},clip, width=0.47\textwidth]{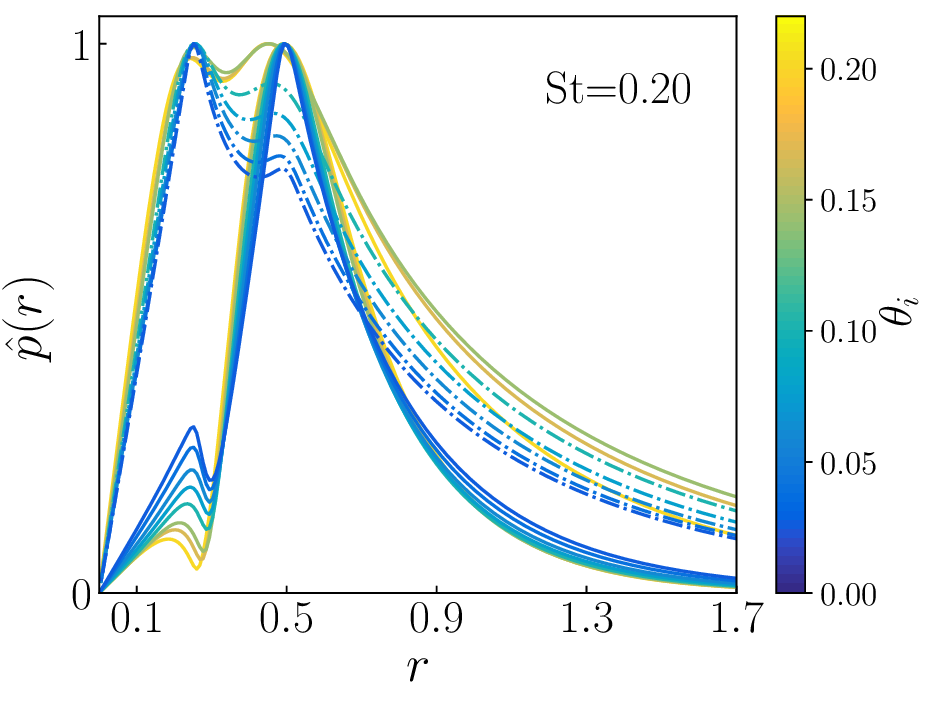}\\
  \includegraphics[trim={0cm 0cm 0cm 0cm},clip, width=0.47\textwidth]{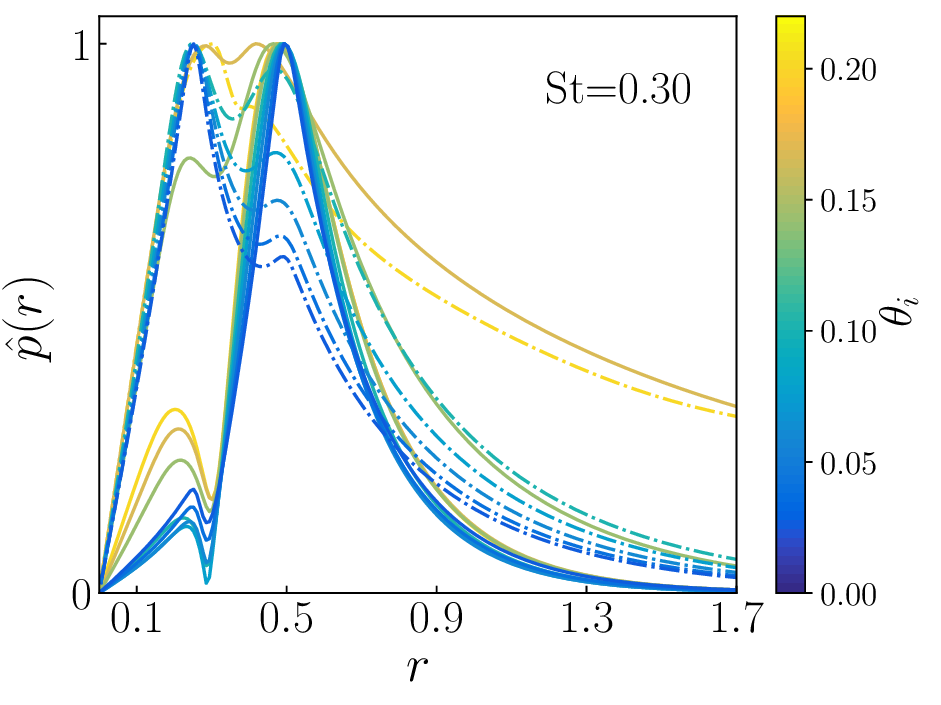}
  \includegraphics[trim={0cm 0cm 0cm 0cm},clip, width=0.47\textwidth]{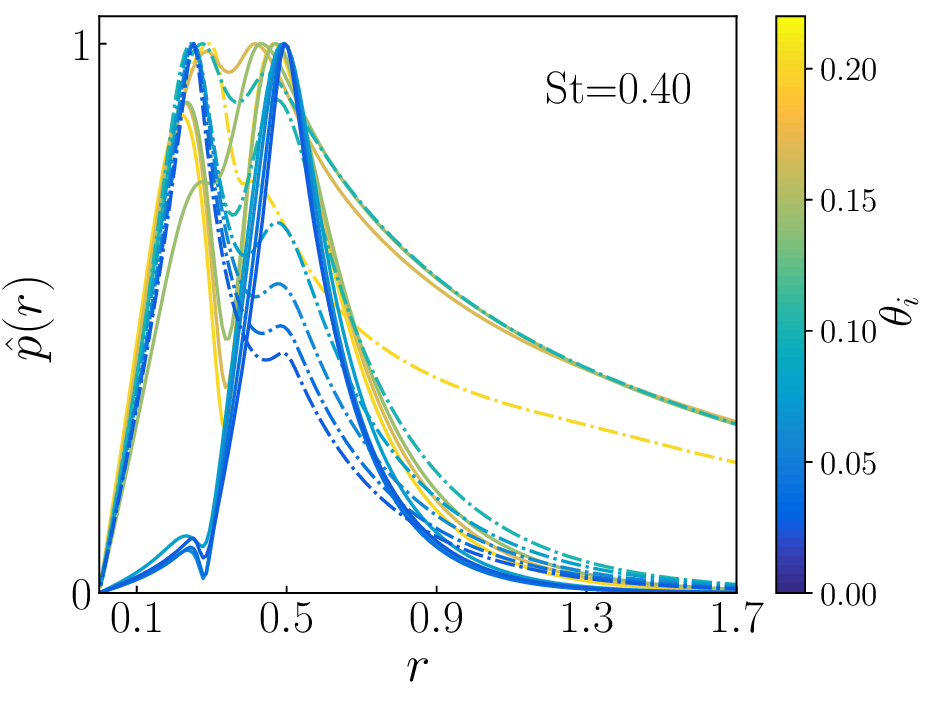}\\
  \caption{Eigenfunctions associated with the inner (dashed) and outer (plain) K-H modes, normalized
    with respect to their maximum values, for $St=0.1$ (top left), $St=0.2$ (top right), $St=0.3$
    (bottom left) and $St=0.4$ (bottom right).}
  \label{fig:kh_func_ML}
\end{figure}

The associated spatial profiles of the KH modes mentioned above are presented in Figure
\ref{fig:kh_func_ML} for different Strouhal numbers and varying mixing layer thicknesses. We observe
similar results to the previous study: for the thinnest mixing layer, the inner and outer KH modes
exhibit maximum pressure fluctuations at the locations where the base flow exhibits the maximum levels of velocity gradients. For all frequencies, thickening the mixing layer tends to spread the spatial support of the inner modes towards the outer mixing layer by increasing the pressure amplitude near the jet boundary. At the highest frequencies studied here, the signature of the outer KH modes also spreads onto the inner mixing layer. As indicated earlier by their relative growth rates, the entanglement of spatial support of the KH modes with increasing mixing layer thickness might make it difficult to distinguish their relative signatures in the fluctuating pressure field. Nonetheless, as mentioned in the previous section, the inner mixing layer KH mode appears to have a slower radial decay compared to the outer one. This supports the findings of \cite{bakulu2021jet}, which show that at the resonance frequency, the downstream energy-carrying mode was supported by the inner mixing layer. The wide radial extent of the eigenfunction, as shown here, also suggests that the inner KH mode can exchange energy with guided-jet modes or acoustic modes, whose support can be located outside the jet.

\subsubsection{Guided jet modes}
\begin{figure}
  \centering
  \includegraphics[trim={0cm 0cm 0cm 0cm},clip, width=0.8\textwidth]{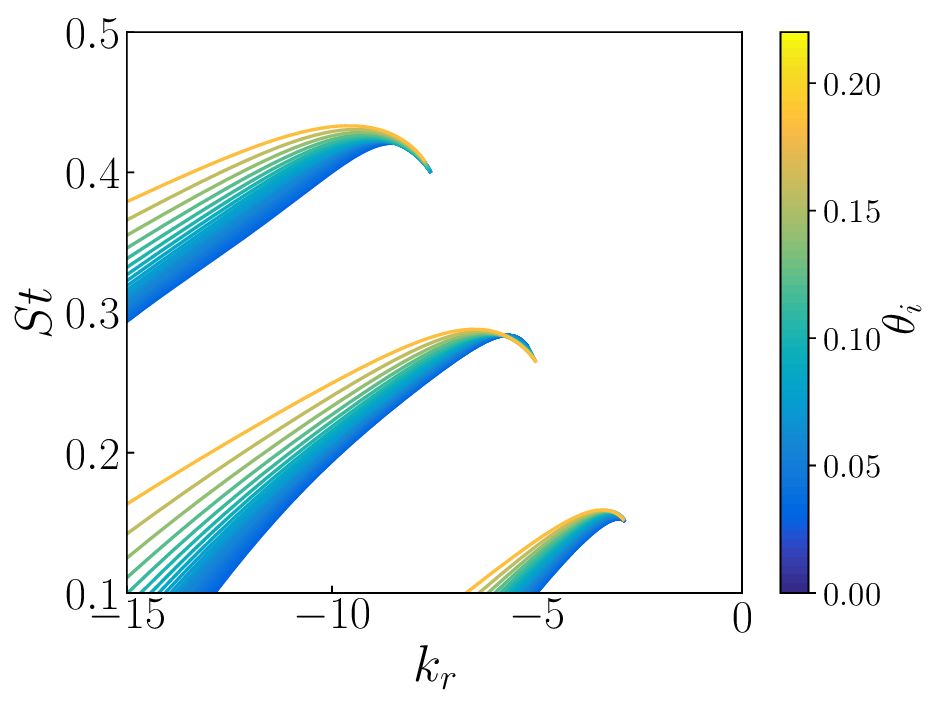}
  \caption{Guided jet modes dispersion relation for varying mixing layer thicknesses.}
  \label{fig:disp_punk_Thick}
\end{figure}

The dispersion relation of the guided jet modes for varying mixing layer thicknesses is now presented in figure \ref{fig:disp_punk_Thick}. Unlike the dispersion relation of K-H modes, the dispersion relation of GJM is only little affected by thicknesses of mixing layers. There is a slight variation in the domain of existence of the upstream-traveling modes between the branch and saddle points, as pointed out by \cite{towne2017acoustic}. The primary effect of thickening the mixing layers is on the group velocity ($\partial \omega/\partial k$), with thicker mixing layers leading to lower group velocities, as one would expect.

\begin{figure}
  \centering
  \includegraphics[trim={0cm 0cm 0cm 0cm},clip, width=0.47\textwidth]{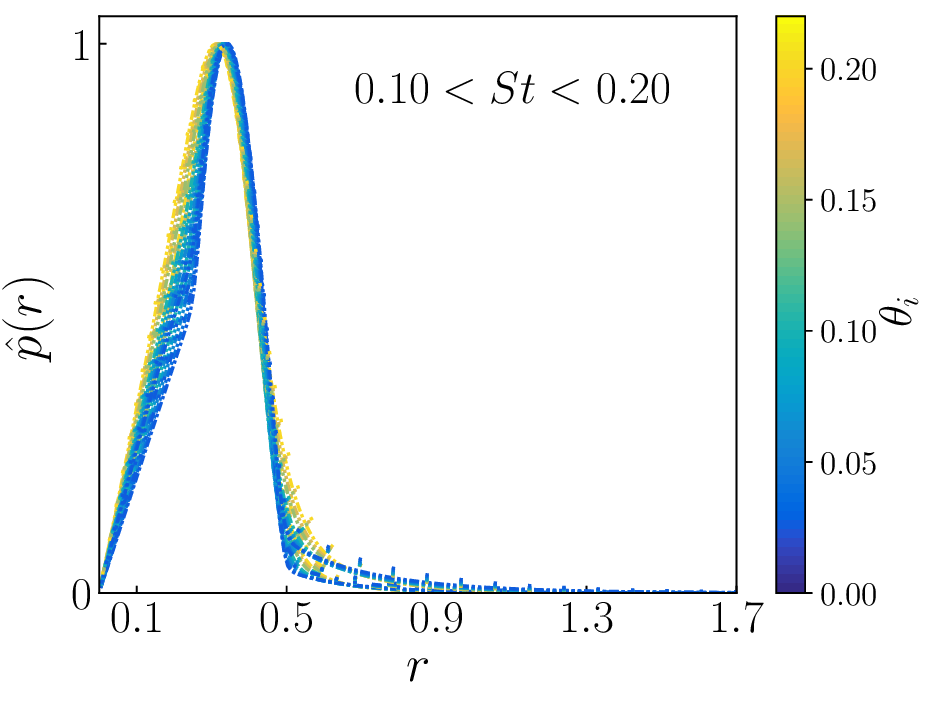}
  \includegraphics[trim={0cm 0cm 0cm 0cm},clip, width=0.47\textwidth]{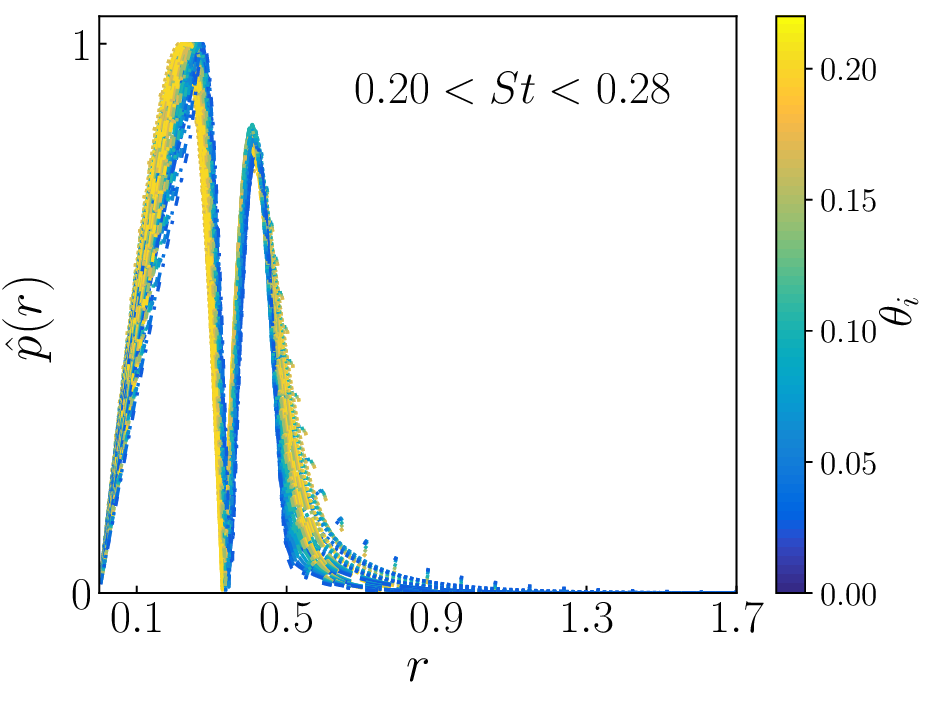}
  \caption{Guided jet modes spatial support for the first radial order (left) and the second one (right). Colors indicate the mixing layer vorticity thickness, and different linestyles are attributed to eigenfunctions of different frequencies.}
  \label{fig:kp_func_ML}
\end{figure}

Figure \ref{fig:kp_func_ML} presents the eigenfunctions of the computed guided jet modes. Frequencies were selected so that modes of the same radial order appear on the same figure. Note that eigenfunctions obtained at the same frequency are plotted with the same linestyle. All computed eigenfunctions almost collapse onto the same line, indicating that the spatial support of the guided jet mode is not affected by the thickening of the jet. The guided jet modes are very robust and can still be observed even if the jet is strongly affected by the diffusion of momentum across the stream interfaces.

\subsection{Stability properties of the flow with varying NPR}
\label{sec:stab_NPR}
\begin{figure}
  \centering
  \includegraphics[height=0.22\textheight]{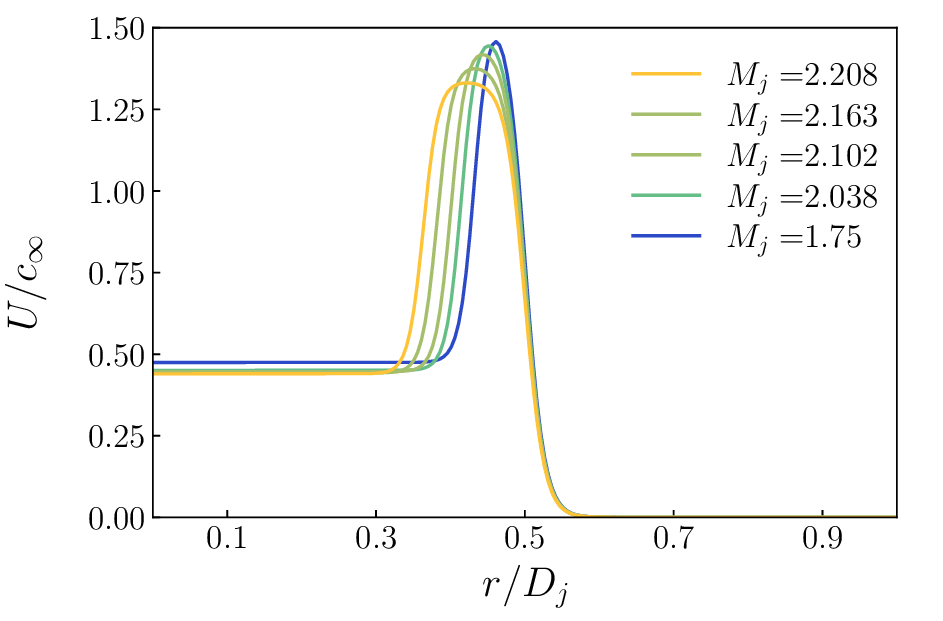}\\
  \caption{Base flows used to study the effect of the expansion regime on the stability characteristics of the flow.}
  \label{fig:baseflow_allMj}
\end{figure}

Using the simple base flow model described in section \ref{sec:mean_flow_model}, we obtained all the
necessary parameters to build simplified base flows for $M_j$ varying in between $1.7$ and $2.2$,
corresponding to expansion regimes where resonances were observed for the TIC nozzle geometry
previously studied \citep{jaunet2017wall}. The aforementioned base flows are presented in figure
\ref{fig:baseflow_allMj} where the reader can see that the position of the inner mixing layer, as
well as both the inner and annular flow velocity, vary with the chosen jet Mach number, their trends
being in agreement with the results of figure \ref{fig:model-Mach}. Regarding the thickness of the
mixing layers, we used that obtained from the numerical results shown in figure \ref{fig:base_flow}
for all cases. In spite of this simplification, it is believed that given the relative
narrow range of jet Mach number considered, the exact mixing layer shape may not drastically change.

\subsubsection{K-H modes}
\begin{figure}
  \centering
  \includegraphics[height=0.22\textheight]{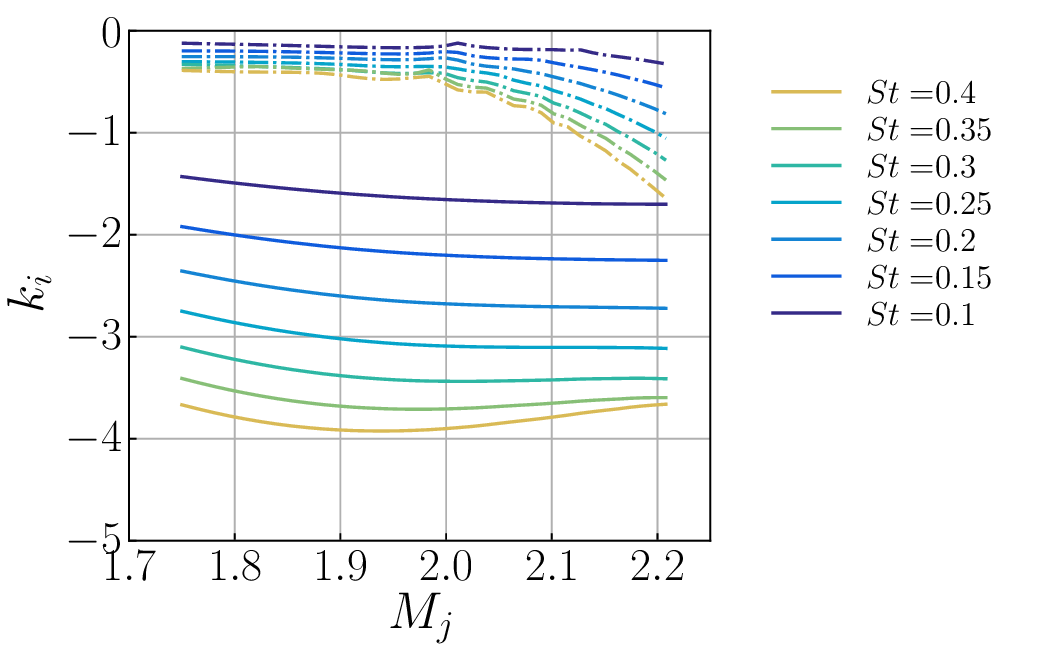}\\
  \caption{Growth rate of both the innner (\dashdotted) and outer (\full) KH modes for varying jet Mach number $M_j$ and Strouhal number.}
  \label{fig:phasespeed_growthrate_allMj}
\end{figure}

The dynamical characteristics of the unstable K-H waves with respect to the base flows presented in
the former paragraph are plotted in figure \ref{fig:phasespeed_growthrate_allMj} for various
frequencies. As expected from the previous results, both KH waves are unstable, the outer one being
the most unstable for all flow regime and frequencies considered. The evolution of the growth rate
of the inner KH wave seems rather independent of the jet Mach number until $M_j = 2.0$. Above this
expansion ratio, it is interesting to notice that the inner KH wave is more and more unstable, for
all the frequencies considered. \vjt{It is worth noting that this value is very close to the value of
$M_j=2.09$ around which resonances were experimentally observed \citep{jaunet2017wall}. The proximity
of these experimental and theoretical values is very striking and still supports the idea that
resonance in TIC nozzle may only be allowed when the inner mixing layer becomes sufficiently
unstable.}

\begin{figure}
  \centering
  \includegraphics[width=0.4\textwidth]{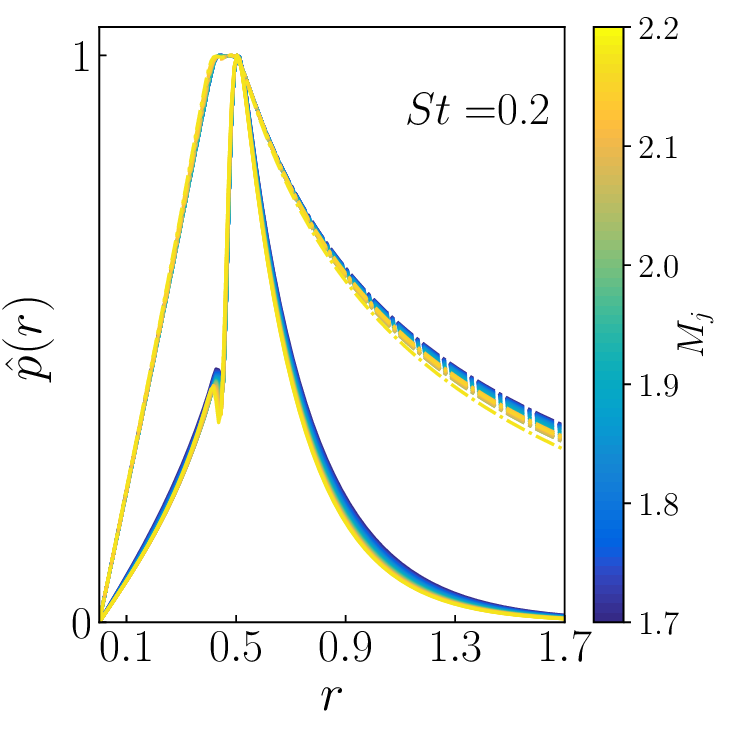}
  \includegraphics[width=0.4\textwidth]{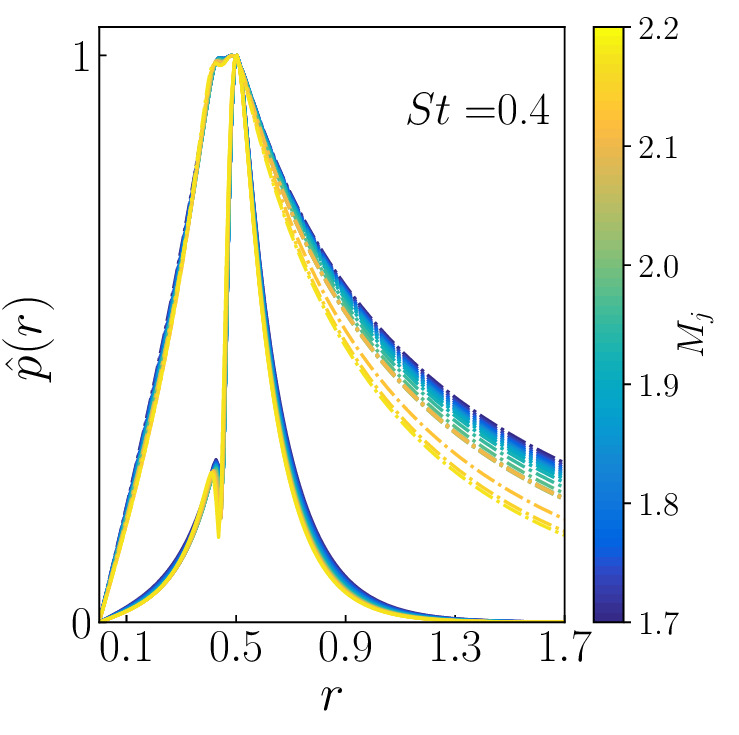}
  \caption{Eigenfunction of both the innner (\dashdotted) and outer (\full) KH modes for varying jet Mach number $Mj$ and Strouhal number.}
  \label{fig:eigenfunction_allMj}
\end{figure}

We report in figure \ref{fig:eigenfunction_allMj} the pressure eigenfunctions of both KH modes. The
outer KH well localized at the location of maximum shear in the outer mixing layer and decays exponentially in the radial direction. For all the cases considered here, the inner KH-mode spatial support is more widely spread across the jet and shows a slower radial decay that the outer one. Overall, the KH waves computed in those cases show the same characteristics than previously. The reader must notice, however, that the spatial support of the KH waves computed does not depend on the Strouhal number nor the jet Mach number.

\subsubsection{Guided jet modes}
\begin{figure}
  \centering
  \includegraphics[height=0.33\textheight]{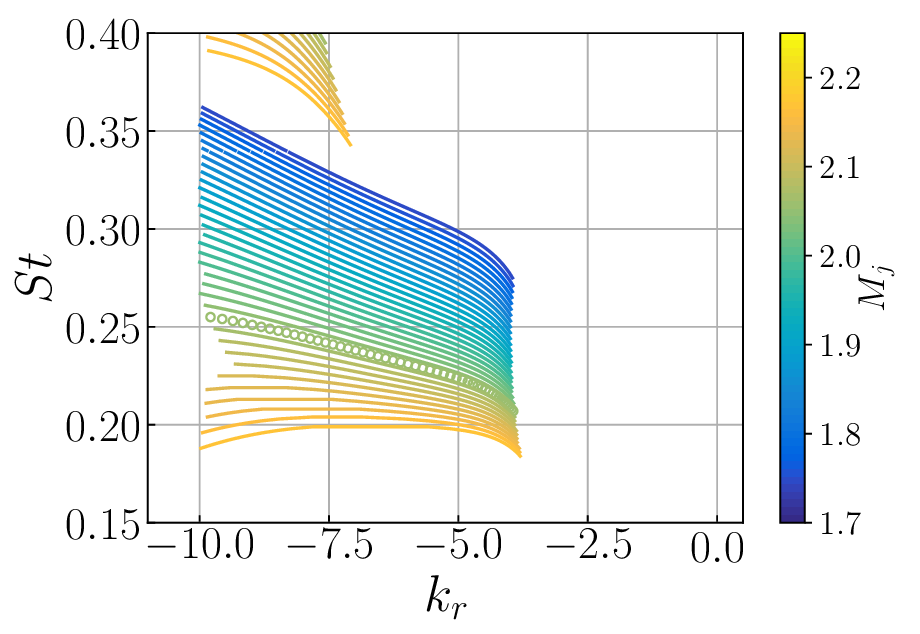}
  \caption{Dispersion relation of the GJM for varying nozzle expansion ratio. The circles represents
  the dispertion relation for $M_j = 2.09$.}
  \label{fig:disprel_allMj}
\end{figure}

As far as guided jet modes is concerned, the expansion ratio of the nozzle plays a significant role
in their dispersion relation, as can be seen in figure \ref{fig:disprel_allMj}. At low $M_j$, the
slope of the dispersion relation is negative and do not show any extrema, hence only upstream
propagating guided jet modes seems supported by the flow field. As depicted earlier in the
study, this is an effect of the thinning of the annular jet providing the modes a behavior
encountered in subsonic jet \citep{jordan2018jet,towne2017acoustic}. As soon as the overall speed of
the jet increases, the GJM dispersion relation deforms towards what is expected in supersonic cases
where the dispersion relation is constituted by both downstream- ($\frac{\partial \omega}{\partial k}>0$)
and upstream-propagating ($\frac{\partial \omega}{\partial k}<0$) branches. In
any case, the results show that for all expansion ratio, the flow supports an upstream-propagating
wave making resonance loops possible.

\begin{figure}
  \centering
  \includegraphics[width=0.4\textwidth]{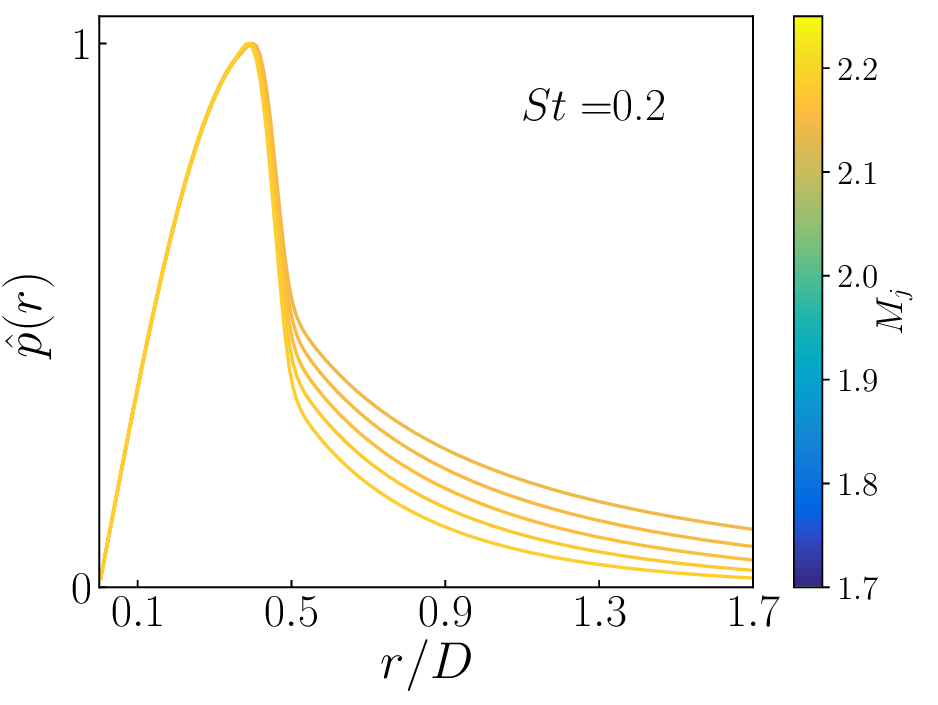}
  \includegraphics[width=0.4\textwidth]{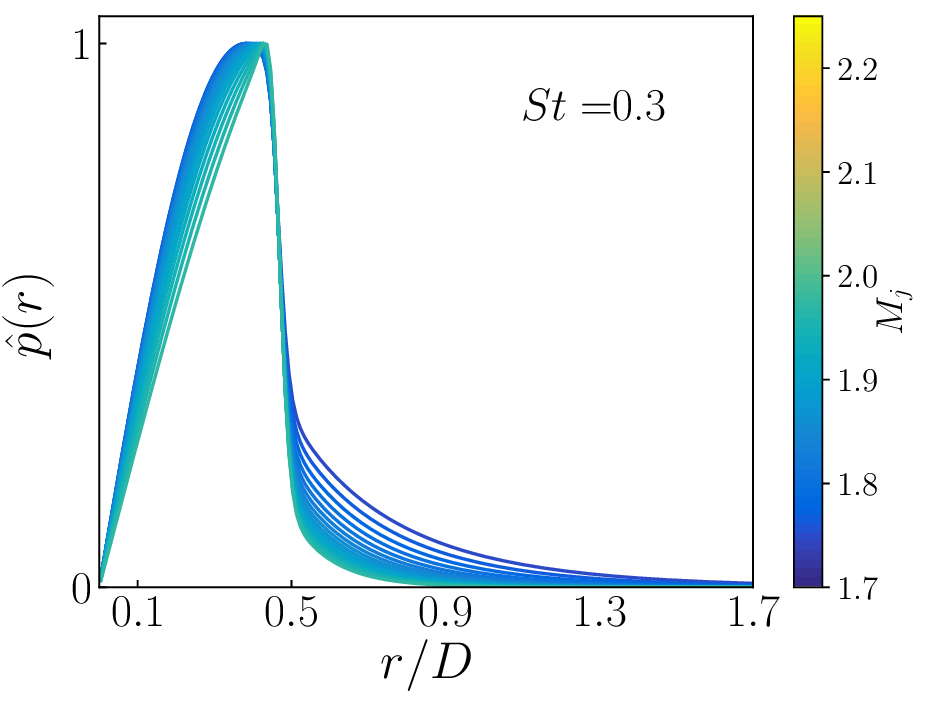}
  \caption{Eigenfunction of the GJM found at $St=0.2$ (left) and $St=0.3$ (Right).}
  \label{fig:eigenfunction_allMj_GJM}
\end{figure}

\subsubsection{Discussion}
According to figure \ref{fig:disprel_allMj}, the dynamical model does not show any upstream
propagating waves for $M_j=2.09$ at $St=0.2$, although a resonance was
experimentally and numerically observed \citep{jaunet2017wall,bakulu2021jet}. This means that it
wouldn't be possible with the current base flow and finite thickness models to predict a resonance
at the correct Mach and Strouhal numbers in over-expanded flows, with a similar mechanism as
depicted for screech like resonance. This is mainly due the numerous
hypothesis and simplification that are embedded in these models.\\
The sensitivity of the waves
dispersion relation and stability characteristics to the base flow, shown in the previous section,
indicates that a more precise definition of the base flow would be necessary if a precise resonance
prediction is needed. The reader may be referred to the work of \cite{chow1975mach} or
\cite{li1998mach} for this purpose.\\
\vjt{The use of a parallel, isobaric flow model might also be cause of this inadequation between the model and
  the observations. Indeed, the flow exiting the nozzle is not parallel. For example, as for
  underexpanded jets, the shock cell network modulates the mean flow in the axial direction aand it
  has been shown to be an important feature in screeching
  jets \citep{nogueira2022absolute}}.\\
Despite these limitations, the analysis conducted in this study revealed that the topology of the
flow exiting a convergent divergent nozzle do support the necessary instability waves to produce a
resonance (\textit{i.e.} downstream KH waves and upstream GJM).

\section{Conclusions and perspectives}
\label{sec:conclusions}
The linear dynamics of inviscid annular supersonic jets, similar to those encountered in
the exhaust of a converging diverging nozzle, was explored in this article. The aim was to provide
insights on the physical origins of tonal dynamics, observed in very limited expansion regimes in
experiments \citep{jaunet2017wall,baars2012wall}, and possibly responsible of unsteady side-loads
\citep{bakulu2021jet,martelli2020flowdynamics}. The focus was therefore put on the first azimuthal
Fourier mode of flow fluctuations, the only one responsible for off-axis loads in axisymmetric
nozzles. The dynamical properties of such fluctuations were computed on base flows with varying
inner mixing layer radial position or varying mixing layer thicknesses. Finally, the effect of a
variable expansion regime, \textit{i.e.} increase in NPR as the flow can encounter during the
start-up of an engine, was explored.\\
As expected from the shape of the base flows, two unstable Kelvin-Helmholtz were found in all
cases. The KH wave supported by the outer mixing layer has most of the time been found more unstable than the
inner one. This is coherent with the differences in velocity gradients in the two mixing layers. Each of
the associated eigenfunctions shows a maximum at the mixing layers location, typical of the KH wave.
The outer KH waves eigenfunctions never showed significant changes as function of the variation of
flow parameters. On the contrary, the inner KH eigenfunctions have shown to be more affected by base flow
changes. Their structure can show a secondary peak at the location of the outer mixing layer and,
more importantly, they seem to have a more pronounced signature outside of the jet than the outer
KH wave. The fact that the inner KH wave has support outside of the jet must be noticed as it allows this wave to
exchange energy with external acoustics or upstream travelling waves. It underpins the fact that the
inner mixing layer, emanating from the Mach disk triple point, is a probable source
of instability in the resonance observed in convergent-divergent nozzles, as conjectured in
\cite{jaunet2017wall} and observed numerically by \cite{bakulu2021jet}.\\
The article purposely took interest in describing the guided-jet modes structure and dynamical
characteristics as they have recently shown responsible for the feedback process in jet resonances
(see \cite{edgington2022Unifying,vare2022generation} for recent references). The most striking
result of the current study is the robustness of the guided-jet modes: they were found in all cases
and their eigenfunctions were shown to be very poorly sensitive to the base flow parameters
tested. This is a clear indication that whenever a jet flow is considered one must pay attention to these
waves as they could be involved in the dynamics. In our case, the dispersion relation and the
existence of the upstream-travelling wave depends strongly on the inner mixing layer position inside
the jet but rather poorly with the mixing layer thicknesses The latter results suggesting that they
could be observed rather far downstream in an exhausting jet that diffuses in the surrounding.\\
Finally, a simplified base flow model was derived in order to evaluate the stability properties of
the flow with varying NPR but arbitrarily fixed mixing layer thicknesses. Like the hypothetical base
flows used in the beginning of the study, two Kelvin-Helmholtz modes and numerous guided jet modes could be
identified in the eigenspectra of the analysis, with characteristics found in line with the
previous results. Interestingly, we have observed that the inner KH wave has rather small growth
rate at low Mach number (NPR) and suddenly increases above the Mach number at which resonances was
experimentally observed. Although this can be fortuitous, this is yet another indication that we have
possibly pinpointed the correct waves at play in the resonances in such flows and that the inner mixing
layer is of great importance in the dynamics of such flows.\\
Although the results of this study revealed interesting feature of the dynamics of annular supersonic
jets, we must recall that an important number of simplifications were made in order to be able to
conduct the analysis. These assumptions are very likely the reason why the prediction of resonance
frequency, as was done in the screeching jet case, was not possible. This provides obvious path for improvement
in both the base flow modeling, by considering a more representative control volume
\citep{li1998mach}, the consideration of the influence of the nozzle
walls or \vjt{even the relaxation of the parallel flow assumption by the use of a global stability
analysis}.

\newpage
\section*{Appendix}
\label{appendix}
\subsection{Mass conservation}

Applying mass conservation on a control volume delimited by the nozzle throat and the attached flow,
as presented in grey in figure\ref{fig:control-volume}, we can write:
\begin{eqnarray}
  \rho_* U_* D_c^2 &=&  \rho_i U_i D_i^2 + \rho_a U_a (D_e^2-D_i^2),
\end{eqnarray}
where $(\cdot)_*$ denotes sonic variables at the throat.\\
Using perfect gaz relations and introducing the speed of sound, such that $U = aM = \sqrt{\gamma r
  T}\cdot M$ and recalling that $M_* = 1$, we have :
\begin{eqnarray}
  \frac{P^*}{r T^*} a^* M^* {D_c}^2 &=&  \frac{P_i}{r T_i} a_i M_i D_i^2 + \frac{P_i}{r T_i} a_i M_i (D_e^2-D_i^2)\nonumber\\
  \frac{P^*}{\sqrt{r T^*}}  {D_c}^2 &=&
                                                    \frac{P_i}{\sqrt{r T_i}}    M_i D_i^2
                                                    + \frac{P_a}{\sqrt{r T_a}}  M_a
                                                     (D_e^2-D_i^2)\nonumber
\end{eqnarray}
Then using isentropic relation to link static and total variables, we obtain :
  \begin{eqnarray}
  {\left(1+\frac{\gamma-1}{2}\right)^{-\frac{\gamma+1}{2(\gamma-1)}}}
  \frac{P_{t0}}{\sqrt{T_{t0}}} D_c^2 &=& \frac{P_i}{\sqrt{T_i}} \mathcal{M}_i D_i^2 +
                                           \frac{P_a}{\sqrt{T_a}} \mathcal{M}_a (1-D_i^2)\nonumber
\end{eqnarray}
Choosing $P_a = P_i = P_\infty$, we can
further simplify the mass conservation equation:
\begin{eqnarray}
{\left(1+\frac{\gamma-1}{2}\right)^{-\frac{\gamma+1}{2(\gamma-1)}}}
  \frac{P_{t0}}{P_\infty} {D_c^2}
  &=& \frac{\sqrt{T_{t0}}}{\sqrt{T_i}} \mathcal{M}_i \frac{D_i^2} +
      \frac{\sqrt{T_{t0}}}{\sqrt{T_a}} \mathcal{M}_a \frac{(D_e^2-D_i^2)}\nonumber\\
{\left(1+\frac{\gamma-1}{2}\right)^{-\frac{\gamma+1}{2(\gamma-1)}}}
  \frac{P_{t0}}{P_\infty} {D_c^2}
    &=&  {D_i^2} { \mathcal{M}_i \sqrt{1+\frac{\gamma-1}{2}\mathcal{M}_i^2}}\nonumber\\
    & &+ {(D_e^2-D_i^2)} { \mathcal{M}_a
        \sqrt{1+\frac{\gamma-1}{2}\mathcal{M}_a^2}},\nonumber
\end{eqnarray}
which can be written in the more compact form presented in the document :
\begin{eqnarray}
\beta_m \frac{D_c^2}{D_e^2}   &=&  (\mu_i - \mu_a) \frac{D_i^2}{D_e^2}  +  \mu_a
\end{eqnarray}
where:
\begin{eqnarray}
  \mu_{i,a}& =&  { \mathcal{M}_{i,a} \sqrt{1+\frac{\gamma-1}{2}\mathcal{M}_{i,a}^2}} \nonumber\\
  \beta_m  &=& {\left(1+\frac{\gamma-1}{2}\right)^{-\frac{\gamma+1}{2(\gamma-1)}}}
  \frac{P_{t0}}{P_\infty}.\nonumber
\end{eqnarray}

\subsection{Momentum conservation}
Now considering the momentum conservation along the axial direction on the same control volume, and neglecting body and viscous forces, leads to:
\begin{eqnarray}
   \gamma P_* {\mathcal{M}_*}^2 D_c^2 + P_* D_c^2
   & = & \gamma P_i \mathcal{M}_i^2 D_i^2 + P_i D_i^2
         + \gamma P_a \mathcal{M}_a^2 (D_e^2 - D_i^2)\nonumber\\
     & &    + P_a (D_e^2 - D_i^2) - F_w \nonumber
\end{eqnarray}
where $F_w$ represents the pressure forces acting on the nozzle wall. Using isentropic relations,
isobaric assumption (as for the mass conservation) and rearanging provides:
\begin{eqnarray}
  (1+\gamma)\left(1+\frac{\gamma-1}{2}\right)^{\frac{-\gamma}{\gamma-1}}
  \frac{P_{t0}}{P_\infty} D_c^2  - \frac{F_w}{P_\infty}
   & = & D_i^2 (\gamma \mathcal{M}_i^2 + 1) + D_e^2(\gamma \mathcal{M}_a^2 + 1) - D_i^2 (\gamma
         \mathcal{M}_a^2 + 1),  \nonumber
\end{eqnarray}
which can be written in the more compact form:
\begin{eqnarray}
  {\beta_q} \frac{D_c^2}{D_e^2} - \frac{F_w}{P_\infty}
   & = & \gamma \frac{D_i^2} {D_e^2} \left( \mathcal{M}_i^2 - \mathcal{M}_a^2 \right) + \left(\gamma \mathcal{M}_a^2 + 1\right),
\end{eqnarray}
where:
\begin{eqnarray}
   \beta_q &=& (1+\gamma)\left(1+\frac{\gamma-1}{2}\right)^{\frac{-\gamma}{\gamma-1}}
   \frac{P_{t0}}{P_\infty},\nonumber
\end{eqnarray}
From the
latter equation we can get an analytical expression for $D_i^2$:
\begin{eqnarray}
    D_i^2 &=& D_e^2\frac{1}{\gamma \left(\mathcal{M}_i^2 -
              \mathcal{M}_a^2\right)}\left[ {\beta_q} D_c^2
              - \frac{F_w}{P_\infty} - \left( \gamma \mathcal{M}_a^2 +1 \right)\right] \nonumber,
\end{eqnarray}
which can be plugged into the mass conservation equation leading to the equation for the annular
Mach number presented in the document:
\begin{eqnarray}
  {\frac{\mu_i - \mu_a}{\gamma \left(\mathcal{M}_i^2 - \mathcal{M}_a^2\right)}
  \left[ {\beta_q}{D_c^2} - \frac{F_w}{P_\infty}
  - \left( \gamma \mathcal{M}_a^2 + 1 \right)\right] }  +
  \mu_a - \beta_m    &=& 0
                                 \label{eqn:annulaD_Mach}
\end{eqnarray}
where:
\begin{eqnarray}
  \mu_{i,a} & =&  { \mathcal{M}_{i,a} \sqrt{1+\frac{\gamma-1}{2}\mathcal{M}_{i,a}^2}}\nonumber\\
  \beta_m  & =& {\left(1+\frac{\gamma-1}{2}\right)^{-\frac{\gamma+1}{2(\gamma-1)}}}
  \frac{P_{t0}}{P_\infty} \nonumber\\
  \beta_q  & =& (1+\gamma)\left(1+\frac{\gamma-1}{2}\right)^{\frac{-\gamma}{\gamma-1}}
  \frac{P_{t0}}{P_\infty}. \nonumber
\end{eqnarray}



\section*{Declaration of Interests}
The authors report no conflict of interest.

\bibliographystyle{jfm}
\bibliography{biblio}

\end{document}